\documentclass[12pt]{article}
\usepackage{amsmath}
\usepackage{graphicx}
\usepackage{enumerate}
\usepackage{natbib}
\usepackage{url} 

\usepackage[T1]{fontenc}
\usepackage{babel}
\usepackage{authblk}
\usepackage{caption}
\usepackage{subcaption}

\usepackage{siunitx}
\usepackage{placeins}
\usepackage{algorithm}
\usepackage{algpseudocode}
\usepackage{bm}
\usepackage{keyval}
\usepackage{setspace}

\newcommand{\blind}{1}

\newcommand{\map}{\mathrm{\tiny MAP}}
\newcommand{\Modbig}{\operatorname{Mod}}
\newcommand{\Argbig}{\operatorname{Arg}}
\newcommand{\ft}{\mathcal{F}}
\newcommand{\iid}{\emph{i.i.d.\,}}
\newcommand{\xval}{2.8}

\newcommand{\yval}{0.28}
\newcommand{\yvalb}{0.45}

\algnewcommand\algorithmicforeach{\textbf{for each}}
\algdef{S}[FOR]{ForEach}[1]{\algorithmicforeach\ #1\ \algorithmicdo}

\begin{document}

\def\spacingset#1{\renewcommand{\baselinestretch}%
{#1}\small\normalsize} \spacingset{1}


\if1\blind
{
  \title{\bf Bayesian Image Analysis in Fourier Space}
  \author{John Kornak\thanks{
    The authors gratefully acknowledge \textit{National Institute of Biomedical Imaging and Bioengineering of the National Institutes of Health award number R01EB022055}}\hspace{.2cm}\\
    Department of Epidemiology and Biostatistics, University of California, San Francisco, USA\\
    and \\
    Karl Young \\
    University of California, San Francisco, USA (Retired)\\
    and \\
    Eric Friedman \\
    International Computer Science Institute, Berkeley, USA
    }
  \maketitle
} \fi

\if0\blind
{
  \bigskip
  \bigskip
  \bigskip
  \begin{center}
    {\LARGE\bf Bayesian Image Analysis in Fourier Space}
\end{center}
  \medskip
} \fi

\bigskip

\begin{abstract}
Bayesian image analysis has played a large role over the last 40+~years in solving problems in image noise-reduction, de-blurring, feature enhancement, and object detection. However, these problems can be complex and lead to computational difficulties, due to the modeled interdependence between spatial locations. The Bayesian image analysis in Fourier space (BIFS) approach proposed here reformulates the conventional Bayesian image analysis paradigm for continuous valued images as a large set of independent (but heterogeneous) processes over Fourier space. The original high-dimensional estimation problem in image space is thereby broken down into (trivially parallelizable) independent one-dimensional problems in Fourier space. The BIFS approach leads to easy model specification with fast and direct computation, a wide range of possible prior characteristics, easy modeling of isotropy into the prior, and models that are effectively invariant to changes in image resolution.
\end{abstract}

\noindent%
\vfill

\newpage
\spacingset{1} 
\section{Introduction}
\label{intro}

Bayesian image analysis models provide a solution for improving image quality in image reconstruction/enhancement problems by incorporating \emph{a~priori} expectations of image characteristics along with a model for image noise, i.e., for the image degradation process~\citep{winkler1995image}.
However, conventional Bayesian image analysis models, defined in the space of conventional images (hereafter referred to as ``image space'') can be limited in practice because they can be difficult to specify and implement (requiring problem-specific code) and they can be slow to compute estimates for.
Furthermore, Markov random field (MRF) model priors in conventional Bayesian image analysis (as commonly used for the type of problem discussed here) are not invariant to changes in image resolution and are generally difficult to specify with isotropic autocovariance \citep{tjelmeland1998markov,rue2002fitting} . 

Our approach to overcoming the difficulties and limitations of the conventional Bayesian image analysis paradigm, is to move the problem to the Fourier domain and reformulate in terms of spatial frequencies. Spatially correlated prior distributions (priors) that are difficult to model and compute in conventional image space, can be more easily modeled via a set of independent priors across locations in Fourier space (at least for images where pixel intensities are on a continuous scale). 

In this Bayesian image analysis in Fourier space (BIFS) formulation, a prior is specified for the signal at each Fourier space location (i.e. at each spatial frequency) and \emph{Parameter functions} are specified to define the values of the parameters in the prior distribution across Fourier space locations The original high-dimensional problem in image space is thereby broken down into a set of one-dimensional problems, leading to easier specification and implementation, and faster computation that is additionally trivially parallelizable. The fast computation coupled with trivial parallelization has the potential to expand the role of Bayesian image analysis to big data problems. Furthermore, the BIFS approach carries with it numerous useful properties, including easy specification of isotropy and consistency in priors across differing image resolutions for the same field of view.

Note that the BIFS approach is distinct from shrinkage prior methods that have been used previously in Fourier, wavelet or other basis set prior specifications (e.g. ~\citet{olshausen1996emergence,levin2007user}). BIFS does not seek to smooth by generating a sparse representation in the transformed space through simple thresholding with the hope that it provides \emph{a~priori} desired spatial characteristics. Rather, the goal of BIFS is to fully specify the prior distribution over Fourier space in the same spirit as Markov random field or other low-level spatial priors in Bayesian image analysis.  

\subsection{Bayesian image analysis}

The general image analysis problem can be described as follows: Consider a true (unobserved) ideal image $x$, that has been degraded by some "noise" process to give the observed image $y$. The goal is to use a prior that can lead to an optimal estimate (or reconstruction) of the undegraded image $x$; alternatively the goal of the prior may be to enhance specific features of $x$, e.g. to enhance tumors in a brain MRI.

\subsection{Conventional Bayesian image analysis:} 

The Bayesian image analysis paradigm incorporates \emph{a~priori} desired spatial characteristics of the reconstructed image via a prior distribution (``the prior'') for the ``true'' image $x$: $\pi(x)$; and the noise degradation process via the likelihood: $\pi(y|x)$. The prior and likelihood are combined via Bayes' Theorem to give the posterior: $\pi(x|y) \propto \pi(y|x) \pi(x)$ from which an estimate of $x$ can be extracted, e.g., the ubiquitous \emph{maximum a posteriori} (MAP) solution obtained by determining the image associated with the mode of the joint posterior distribution. 

The most common choice for the prior in conventional (low-level) Bayesian image analysis is a Markov random field (MRF) model~\citep{geman1984stochastic,besag1989digital,besag1991bayesian}. MRF priors are used for imposing expected \emph{contextual information} to an image such as spatial smoothness, textural information (small-scale pattern repetition), edge configurations (patterns of locations of boundaries with large intensity differences) etc.  MRF methods provide improvement over deterministic filtering methods by \emph{probabilistically} interacting with the data to smooth, clean, or enhance images, by weighting information from the data (via the likelihood) with the MRF prior to form the posterior. 

\subsection{Fourier space methods for image analysis} 

There is considerable literature on methods for representing processes in terms of basis set representations, see e.g., in the field of functional analysis~\citep{morris2014functional}. Fourier/wavelet basis set shrinkage/thresholding based methods have seen multiple applications to the field of image analysis (including from a Bayesian perspective through L1/L2 regularization). In particular, methods for image processing have been developed using Fourier, wavelet and other bases sets e.g.,~\cite{donoho2002beamlets,Chang2000adaptive,levin2007user,pavlicova2008detecting,li2014bayesian}. The sparse representation in the transformed space, e.g. Fourier, leads to noise reduction back in image space. However, in contrast to the BIFS approach that we develop here, there is no explicit \emph{a~priori} model for expected structure in the true image, other than that the true image might be well represented by a small subset of the basis functions. The BIFS paradigm presented here provides a comprehensive approach to characterizing image priors by modeling specific priors at all Fourier space locations.

Fourier representations of Gaussian Markov random fields (GMRFs) have been used in order to generate fast simulations when the GMRF neighborhood structure can be represented by a block-circulant matrix, i.e. such that the GMRF can be considered as wrapped on a 2D torus:~see~Ch.~2.6 of~\citet{rue2005gaussian}. We will explore the relationship between this Fourier space representation and a special case of the Bayesian image analysis in Fourier space approach we are proposing in Section~\ref{MRImatch}.

For the remainder of the paper, in Section \ref{secmodel} we describe the BIFS modeling framework and provide example models and implementations. In Section \ref{secproperties} we examine some properties of working with BIFS. In Section \ref{secapprox} we consider how to approximate convention Bayesian image analysis models with BIFS. For Section \ref{secdatadriven} we examine a data-driven approach to implementing BIFS and finally in Section \ref{secconclude} we provide some discussion and conclusions.

\section{BIFS Modeling Framework} \label{secmodel}

Consider $x$ to be the true (or idealized/enhanced) image that we wish to recover from a somehow degraded image dataset $y$. 
Instead of the conventional Bayesian image analysis approach of generating prior and likelihood models for the true image $x$ based on image data $y$ directly in terms of pixel values, we formulate the models via their discrete Fourier transform representations: $\ft x$ and $\ft y$. Using Bayes' Theorem, the posterior, $\pi(\ft x | \ft y)$, is then, 
\begin{equation} 
\pi(\ft x | \ft y) \propto \pi(\ft y | \ft x) \pi(\ft x) \enspace.   
\end{equation}

The key aspect of the BIFS formulation that leads to its useful properties of easy specification and computational speed, is that we specify both the prior and likelihood (and therefore the posterior) as a set of independent processes over Fourier space locations. The BIFS prior is able to induce spatial correlation in image space by specifying the parameters of the prior distributions such that they change in a systematic fashion over Fourier space; 
 independent processes in Fourier space are thereby transformed into spatially correlated processes in image space~\citep{zeger1985exploring,lange1997non,peligrad2006central}. Heuristically, the realized signal at each position in Fourier space corresponds to a spatially correlated process in image space (at one particular spatial frequency). In general therefore, linear combinations of these spatially correlated signals (such as that given by the discrete Fourier transform) will also lead to a spatially correlated process in image space. This independence-based specification over Fourier space can be contrasted with the conventional Bayesian image analysis approach of using Markov random field (MRF) priors, where neighborhood properties are used to induce correlation patterns across pixels via joint or conditional distributional specifications \citep{geman1984stochastic,besag1989digital}. In certain instances MRF models exactly correspond to uncorrelated processes in Fourier space (see Section \ref{MRImatch}).

When specifying a BIFS prior for a spatially correlated process in image space via a set of independent processes across Fourier space, the full conditional posterior at a Fourier space location
$k= (k_x,k_y) \in [-\pi, \pi)^2$, or for volumetric data $(k_x,k_y,k_z) \in [-\pi, \pi)^3$, trivially only depends on the prior and likelihood at that same Fourier space location $k$, i.e., 
\begin{equation}
\pi(\ft x_k | \ft y)  = \pi(\ft x|  \ft y_k) \propto \pi(\ft y_k| \ft x_k) \pi(\ft x_k) \enspace,
\end{equation}  
where we use $\ft x_k$ as shorthand for $(\ft x)_k$. The joint posterior density for the image is then 
\begin{equation}
\pi(\ft x|  \ft y) \propto \prod_{k \in K} \pi(\ft y_k| \ft x_k) \pi(\ft x_k) \enspace,
\end{equation}
where $K$ is the set of all Fourier space point locations in the (discrete) Fourier transform of the image. Note that for our purposes we index Fourier space along direction $v \in \{x,y,z\}$ by $\{-N_v/2,\ldots,0,1,\ldots,N_v/2 - 1\}$, rather than the common alternative of, $\{0,\ldots,N_v-1$\}; therefore, the center point $(0,0)$ corresponds to the zero frequency position of Fourier space. Furthermore, since most images are in practice real-valued, the Fourier transform must be conjugate (Hermitian) symmetric on the plane (or volume if 3D). A real-valued image output is ensured by specifying the prior to be conjugate symmetric with respects to Fourier space coordinates (see \citet{liang2000principles}, pp. 31 and 322).  Therefore, for real-valued images, the BIFS posterior only needs to be evaluated over half of Fourier space (and points on the line $x=0$ if taking half-plane in the $y$-direction or conversely $y=0$ for half-plane in the $x$-direction) and the remainder is obtained by conjugate reflection.

In defining priors as a process over Fourier space we are restricting the space of possible priors to stationary processes that are toroidally wrapped around at the edges, analogous to MRFs with neighborhood structure wrapped on the torus. Note that these priors can be non-stationary with respect in the specific sense of not identifying the level of the overall mean by placing an improper uniform prior at k-space point $(0,0)$ for the modulus, i.e., leading to models with the same property as the intrinsic MRFs. In practice, for image analysis problems where the goal is to enhance features, the impact of this restriction to stationarity on the torus primarily affects the edges of an image. However, these effects can be mitigated by expanding the field of view of the data, e.g., by setting pixel values in the expanded boundaries to the overall image mean or to some local neighborhood mean. Furthermore, in many medical imaging applications the area of interest is far from the edges of the field of view with the image boundary corresponding to regions outside of the body and corresponding intensity levels are flat. 

\subsection{The BIFS prior and parameter functions} \label{methods}

A two-step process is used to specify the BIFS prior distribution over Fourier space. 
First, the distributional form of the prior for the signal intensity at each Fourier space location is specified, i.e., $\pi(\ft x_k)$. 
Second, the parameters of each of the priors are specified at each Fourier space location using \emph{parameter functions}. 
These functions specify the parameter values across all Fourier space locations simultaneously, i.e., for some parameter $\alpha_k$ of $\pi(\ft x_k)$ we set $\alpha_k  = f_{\alpha}(k)$. For most problems in practice it is desirable to choose a spatially isotropic prior, which can be induced by specifying $\alpha_k = f_{\alpha}(|k|)$, where $|k| = \sqrt{k_x^2 + k_y^2}$ in 2D  (or $\sqrt{k_x^2+k_y^2+k_z^2}$ in 3D) i.e., such that $f$ only depends on the distance from the origin of Fourier space. In the remainder of this paper, description is given in terms of 2D BIFS, but notational extension and application to 3D volumetric imaging is straightforward.

\begin{figure}
\centering
    \begin{subfigure}{.4\textwidth}
      \centering
       \includegraphics[width=5.5cm,height=2.2cm,trim={0.0cm 6.5cm 0.0cm 3.5cm}, clip=TRUE,page=2]{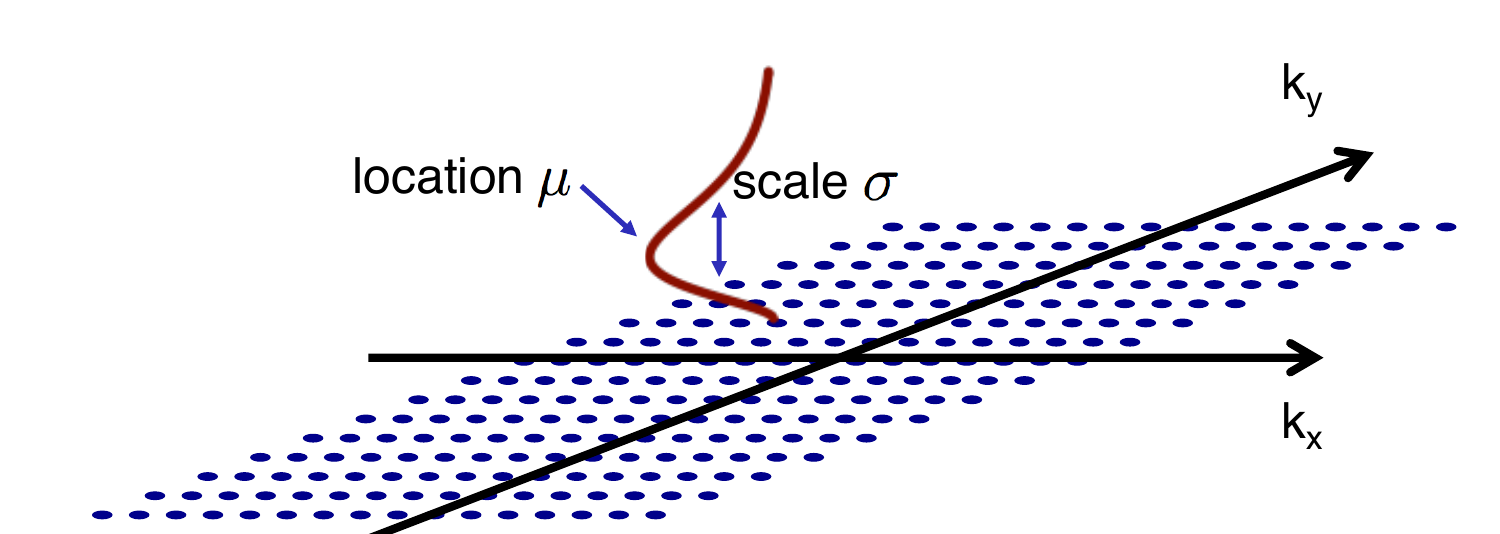}
        \caption{}\label{fig:fig_a}
    \end{subfigure} %
    \begin{subfigure}{.4\textwidth}
      \centering
      \includegraphics[width=5.5cm,height=2.2cm,trim={0.0cm 6.5cm 0.0cm 3.5cm}, clip=TRUE,page=4]{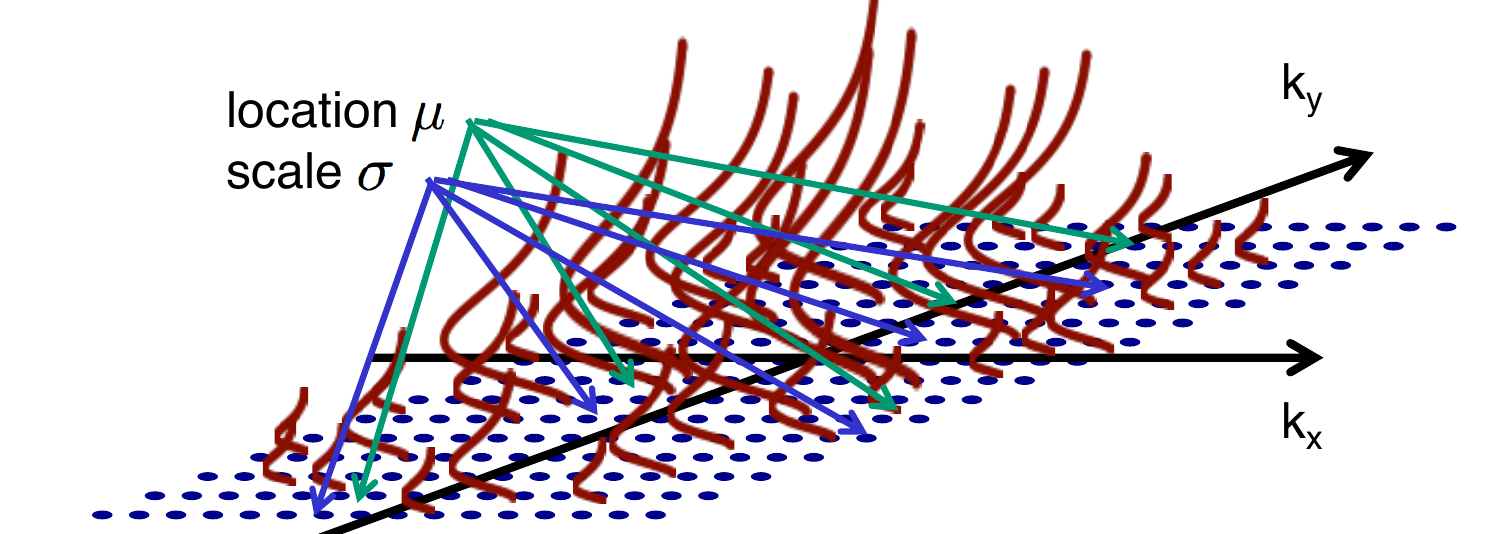}
        \caption{}\label{fig:fig_b}
    \end{subfigure} %
    \begin{subfigure}{.4\textwidth}
      \centering
      \includegraphics[width=5.5cm,height=2.2cm,trim={0.0cm 6.5cm 0.0cm 3.5cm}, clip=TRUE,page=5]{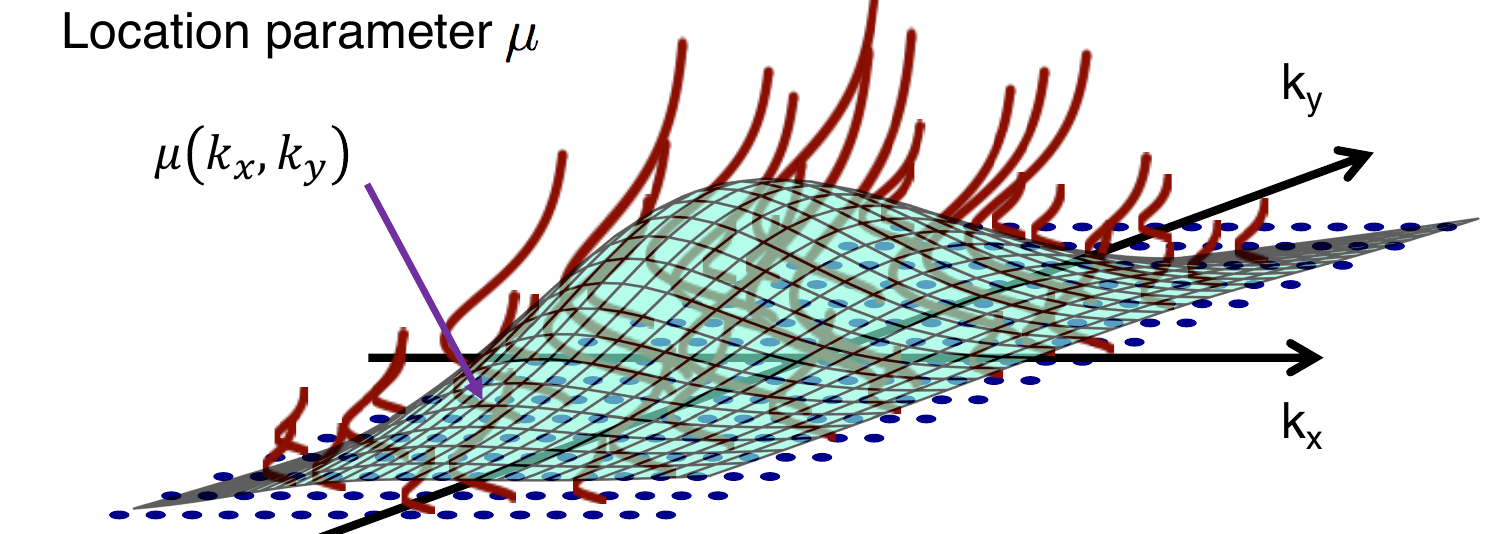}
        \caption{}\label{fig:fig_c}
      \end{subfigure}
      \begin{subfigure}{.4\textwidth}
      \centering
      \includegraphics[width=5.5cm,height=2.2cm,trim={0.0cm 6.5cm 0.0cm 3.5cm}, clip=TRUE,page=8]{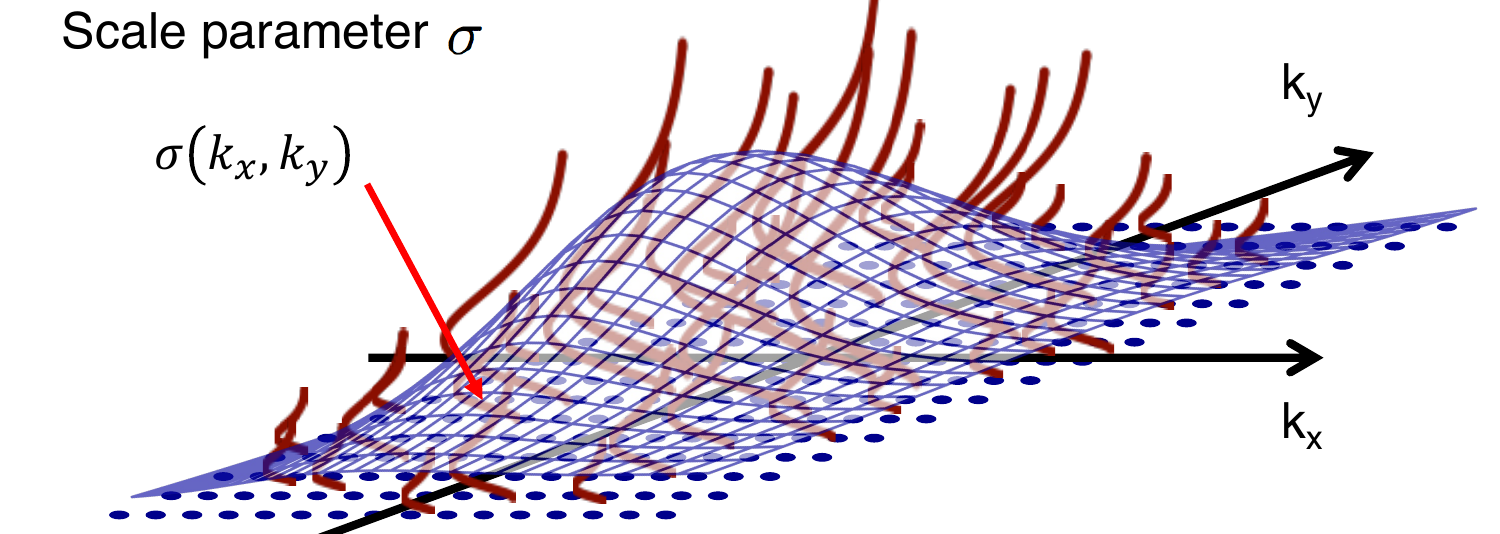}
        \caption{}\label{fig:fig_d}
      \end{subfigure}
\caption{Schematic of parameter function for the signal modulus where the distribution at each Fourier space location requires specification of a location parameter, $\mu$, and scale parameter, $\sigma$. Panel~(\subref{fig:fig_a}) gives the layout of Fourier space with a pdf for a single Fourier space location. Panel~(\subref{fig:fig_b}) extends to show pdfs at multiple Fourier space positions, each of which has a location parameter as described by the parameter function for location shown as a surface mesh in Panel~(\subref{fig:fig_c}) and a scale parameter described by the parameter function for scale in Panel~(\subref{fig:fig_d}). The scale parameter function which is here assumed to be proportional to that of the location parameter, i.e. $\mu(k_x,k_y)=c\sigma(k_x,k_y)$; $c>0$. This can be a practical, but not necessary assumption. \label{parfn} }
\end{figure}

The parameter function traces out the values of parameters for the prior over Fourier space (see illustrations of Figure~\ref{parfn}). In general, the parameter function is multivariate, with one dimension for each parameter of the prior distribution used at each Fourier space location, e.g. Figure~\ref{parfn} provides a schematic for the specification of parameter functions for distributions with two parameters (scale and location). 

Separate priors, and associated parameter functions, are specified for each of the modulus and argument of the complex value at each Fourier space location. Working with the modulus and argument provides a more convenient framework for incorporating prior information at specific Fourier space locations (i.e., specific spatial frequencies) than working with the real and imaginary components. The convenience arises because prior information (e.g., expected characteristics of smoothness, edges, or features of interest) can be directly specified via the modulus of the process, with the argument being treated independently of the modulus. 

Real and imaginary components are more difficult to specify since signal can shift between them through the translation of objects in the corresponding image space; for example, a rigid movement of an object in an otherwise constant intensity image will cause shifts between real and imaginary components (by the Fourier transform shift Theorem) whereas in the modulus/argument specification it will only change the argument of the signal. 

\subsection{Priors and parameter functions for signal modulus}

\subsubsection*{Modulus parameter function forms:}
The parameter functions for the modulus are specified as a set of 2D functions over Fourier space (in terms of $k_x$ and $k_y$): often one for each of location (e.g. mean) and scale (e.g. standard deviation, sd), or/and any other parameters of the prior.
The center of Fourier space, i.e., the $k = (k_x,k_y) = (0,0)$ frequency, is the prior for the overall image intensity mean. At this location, it is reasonable to choose a different pdf than at all other locations. Indeed, the prior at $k = (k_x,k_y) = (0,0)$ can be modeled as improper, e.g. uniform on the real line, leading to an intrinsic, non-stationary prior for the image \citep{kunsch1987intrinsic,besag1991bayesian}.

In general, useful priors across Fourier space are generated by allowing the parameter function for the location parameter to decrease with increasing distance from the center of Fourier space. In our experience, the functional form of the descent from $k = (0,0)$, as a function of distance, has a major impact on the properties of the prior model; much more so than the form of the pdf chosen for each Fourier space location. In addition, specific ranges of frequencies can be accentuated by increasing the parameter function for the location parameter over those frequencies, leading to enhanced spatial frequency bands as a function of distance from the origin. 

\subsubsection*{Modulus prior form:}
We specify non-negative prior distributions at each point in Fourier space. The distribution itself could be allowed to differ in different regions of Fourier space, though we do not pursue that here except for the specific case of allowing for a different distributional form at the $k = (0,0)$ spatial frequency (corresponding to the mean intensity); c.f., intrinsic MRFs. For the purpose of computational expedience it therefore often make sense to choose conjugate priors to the likelihood for the modulus when available. For example, we could use a normal prior for the location parameter of a lognormal likelihood. 

When considering how to define the modulus parameter function for the scale parameter, we find that setting it to be proportional to that for the mean is often a good strategy though the method allows for different strategies wherever warranted.

Mixtures of distributions could also be considered. In particular, for a mixture of two distributions one of them could consist of a probability mass specified at zero. The variation in probability mass for zero at each Fourier space location can itself be specified by a parameter function. For example, one might want to encourage sparsity in the Fourier space representation by increasing the probability of exact zeros the further away a point is from the origin of Fourier space. 

\subsection{Priors and parameter functions for signal argument}

We specify the prior for the argument to reflect \emph{a~priori} ignorance by using an \iid uniform distribution on $[-\pi, \pi)$. Generally, we have limited prior knowledge about the argument of the signal; the argument is related to the relative positioning of objects in the image; moving (shifting) objects around in an image will change the argument of related frequencies. (This could be leveraged for modalities that need to adjust for movement because a rigid body shift manifests as a single additive adjustment to all arguments over Fourier space.) An exception to using uniform priors for the argument can be useful when the prior is being built empirically from a database of images (Section~\ref{simstudy}).

\subsection{BIFS likelihood} 

As for the prior, the BIFS likelihood is modeled separately for the modulus and argument of the signal at each Fourier space location and denoted $\pi(\ft y_k| \ft x_k)$. 

The parameter(s) of the likelihood needs to be provided or estimated for the BIFS algorithm. A straightforward approach to this estimation is to extrapolate any areas of the original image that are known to consist only of noise to an image of equal size to the original image, and then Fourier transform to estimate the corresponding noise distribution in Fourier space.

In practice, not knowing the noise standard deviation is not a major impediment to proceeding with Bayesian image analysis. The noise in the image and the precision in the prior trade off with one another in the Bayesian paradigm and therefore the parameter functions of the prior can be adjusted to produce a desired effect \emph{a posteriori}. This \emph{ad hoc} approach is often applied in practice for MRF prior modeling in Bayesian image analysis; the appropriate setting of hyper-parameter values is a difficult problem and often they are left as parameters to be tuned by the user \citep{sorbye2014scaling}.

\subsection{Posterior estimation}  \label{argMAP}
 Posterior estimation in conventional Bayesian image analysis tends to focus on MAP estimation (i.e., minimizing a $0-1$ loss function) primarily because it is typically the most computationally tractable. In the BIFS formulation the MAP estimate can be efficiently obtained by independently maximizing the posterior distribution at each Fourier space location, i.e, $x_{\map} = \ft^{-1}(\ft x_{\map})$ where $\ft x_{\map} = \{\ft x_{k, \map}, k = 1, \ldots, K \} $ and $ \ft x_{k, \map} = \max_{\ft x_k} \left\{ \pi(\ft x_k | \ft y) \right\} = \max_{\ft x_k} \left\{ \pi(\ft x_k | \ft y_k) \right\} $. This simple estimation approach contrasts with conventional Bayesian image analysis, where even the most computationally convenient MAP estimates typically require iterative computation methods such as conjugate gradients or expectation-maximization. Beyond the MAP estimate, it is straightforward to simulate from the posterior of BIFS models to get mean estimates such as minimum mean squares estimate (MMSE) estimates \citep{winkler1995image} or other summaries of samples from the posterior. The independence of posterior distributions over Fourier space implies that low-dimensional simulations at each Fourier space location is all that is required to generate samples; in contrast to the high-dimensional MCMC simulations usually required for conventional Bayesian image analysis.

Furthermore, when an uninformative uniform prior is used for the argument, and the likelihood is symmetric about the observed argument in the data, the corresponding maximum of the posterior at that Fourier space point is simply the argument of the Fourier transformed data at that point. Therefore, under these conditions, the exact form of the likelihood for the argument is unimportant in obtaining the MAP estimate. 

\begin{algorithm}
\caption{General BIFS implementation} \label{bifsalg}
\begin{algorithmic}
\State Fast Fourier transform (FFT) image data, $y$, into Fourier space, $\ft y$
\State Specify noise distribution/likelihood in Fourier space $\pi(\ft y_k | \ft x_k)$
\State Specify prior distribution form $\pi(\ft x_k)$
\State Specify parameter functions for each of modulus and argument of the signal
\ForEach{$k \in K$, }
\State Obtain $\pi(\ft x | \ft y) \propto \pi(\ft y | \ft x) \pi(\ft x)$
\State Generate posterior estimates/summaries/simulations at each $k$ via MAP or Monte Carlo
\EndFor
\State Inverse FFT posterior estimates/summaries/simulations back to image space.
\end{algorithmic}
\end{algorithm}

\subsection{Modeling Gaussian \iid noise in image space} \label{ricelik}

A common model for the noise in images is to assume that the image intensities are contaminated by \iid~Gaussian noise. The \iid Gaussian noise in image space transforms to complex Gaussian noise in Fourier space with independent real and imaginary components. However, when the real and imaginary components are transformed to modulus and argument, the associated errors are Rayleigh for the modulus and uniform for the argument. 
The corresponding likelihood model for the modulus is the Rician distribution which takes the following form \citep{rice1945mathematical,gudbjartsson1995rician,rowe2004complex,miolane2017template}: 
\begin{equation}
\pi(r|\rho,\sigma) = \frac{r}{\sigma^2} \exp \left(- \frac{r^2 + \rho^2}{2 \sigma^2} \right) I_0 \left( \frac{r \rho}{\sigma^2} \right); \qquad r, \rho, \sigma \ge 0 \label{ricelikform}
\end{equation}
where $I_0(z)$ is the modified Bessel function of the first kind with order zero, $\sigma$ is the standard deviation of the real and imaginary Gaussian noise components in Fourier space (\iid real and imaginary parts), $r$ is the observed modulus of the signal, i.e., $\Modbig (\ft y_k)$ in the BIFS formulation, and $\rho$ is the noise-free modulus, $\Modbig (\ft x_k)$.

Note that in the Rician likelihood, the standard deviation of each of the real and imaginary components is the $\sigma$ parameter of the Rician distribution. Therefore, by obtaining an estimate of the noise level in image space, the $\sigma$ parameter over Fourier space can itself be estimated by dividing the estimated standard deviation of the noise in image space by $4$. \label{constvar}

The corresponding likelihood for the argument is not explicitly required for MAP estimation (see Section~\ref{argMAP}) but for completeness we include here \citep{gudbjartsson1995rician,rowe2004complex}:
\begin{equation}
\pi(\psi|\rho,\theta,\sigma) = \frac{\exp \left(-\frac{\rho^2}{2 \sigma^2}\right)}{2 \pi}\left[1 + \frac{\rho}{\sigma} \cos(\psi - \theta) \exp \left(\frac{\rho^2 \cos^2 (\psi - \theta)}{2 \sigma^2} \right) \int_{r = - \infty}^{\frac{\rho \cos (\psi - \theta)}{\sigma}} \exp \left( - \frac{z^2}{2} \right) \, dz \right] \label{ModPhase}
\end{equation}
where $\psi \in [-\pi, \pi)$ is the observed argument of the signal, $\Argbig (\ft y_k)$, and $\theta \in [-\pi, \pi)$ is the noise-free argument, $\Argbig (\ft x_k)$.

Direct off-the-shelf optimization is problematic for the Rician likelihood. At many Fourier space locations the prior and likelihood can be highly discordant, i.e. the modes of the prior and the likelihood can be very far apart with very little density in between for both distributions. In practice, direct numerical optimization tends to break down in extremely discordant cases. This discordance is not surprising and exists for the same reason that typical conventional image analysis priors are not full representations of prior beliefs for an image, but instead typically only represent expected local characteristics such as smoothness of the image \citep{besag1989digital,green1990bayesian}. \label{discord} 

We therefore propose the following approach to MAP estimation with the Rician likelihood (we drop the $k$ subscript for location in Fourier space to aid clarity). The posterior is $\pi(\rho|r,\sigma) \propto \pi(r|\rho,\sigma) \pi(\rho)$ and we can take logs to simplify, drop constant terms, and find the maximum. Finally, take the second derivative to check that it is concave by bounding the derivative of the Bessel function.

For example, if we assume an exponential prior for $\rho$ i.e. $\pi(\rho) \propto \exp \left(- \frac{\rho}{m} \right)$, take logs and simplify we get
\begin{equation*} 
    \log  \pi(\rho | r, \sigma) = c + \log \left( \frac{r}{\sigma^2} \right) - \frac{r^2+\rho^2}{2\sigma^2} + \log \left[I_0 \left( \frac{r \rho}{\sigma^2} \right) \right] -\frac{\rho}{m}
\end{equation*}
where $c$ is a constant term. Now differentiate w.r.t. $\rho$ and set to $0$ 
\begin{equation*}
    - \frac{\rho}{\sigma^2} - \frac{1}{m} + \frac{r I_1 \left( \frac{r \rho}{\sigma^2} \right)}{\sigma^2 I_0 \left( \frac{r \rho}{\sigma^2} \right)} = 0
\end{equation*}
where $I_0(z)$ is the modified Bessel function of the first kind with order one, and define 
\begin{equation*}
    b(\rho) = \frac{I_1 \left( \frac{r \rho}{\sigma^2} \right)}{I_0 \left( \frac{r \rho}{\sigma^2} \right)}
\end{equation*}
to get
\begin{equation} \label{exprho}
    \rho = r b(\rho) - \frac{\sigma^2}{m}
\end{equation} 
The posterior estimate of $\rho$ can then be estimated quickly through iteration. Start with $\rho_0=r$ and then iterate Equation~\ref{exprho} repeatedly as $\rho_{n+1} = r b(\rho_n) - \frac{\sigma^2}{m}$ to compute the MAP estimate of $\rho$. In practice this requires only a few iterations to get high accuracy. Note that the unconstrained maximum of the function may occur at negative values of $\rho$. In that case the posterior maximum for $\rho$ is set to $0$ because the function is monotonically decreasing to the right of the maximum. 

\textbf{Argument MAP estimate:} Note that for a noise process with a uniformly random argument on the circle (as for \iid noise in real and imaginary components), the likelihood at a Fourier space point has highest density at the argument of the data at that Fourier space location, i.e. $\Argbig(\ft y_k)$. Therefore, since by adopting a uniform prior for the argument on the circle, then the argument corresponding to the maximum of the posterior is also $\Argbig(\ft y_k)$. 

\subsection{Example 1 -- Smoothing/denoising} \label{denoise}

\begin{figure}
\centering
    \begin{subfigure}{\yval \textwidth}
      \centering
       \includegraphics[width=\xval cm,height= \xval cm]{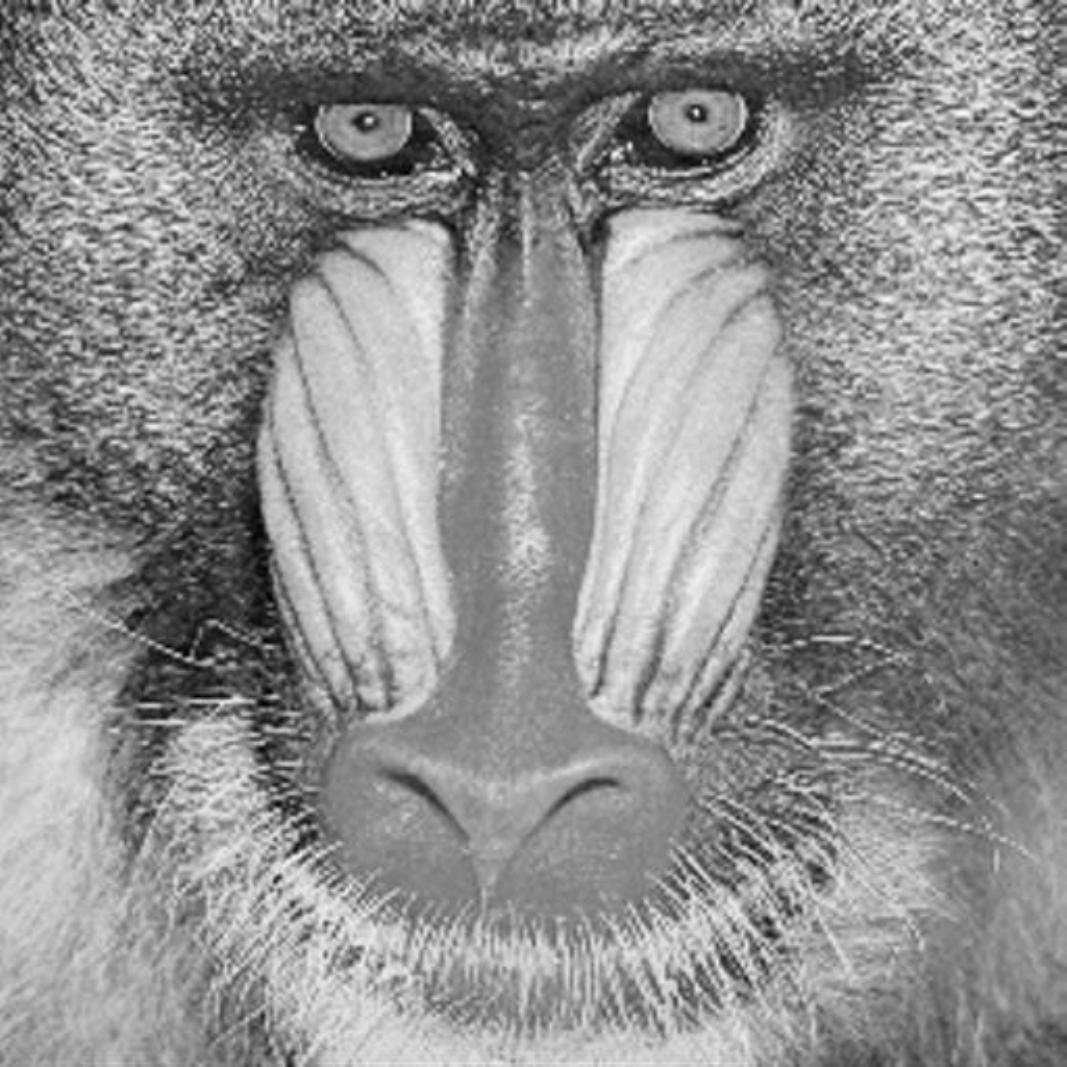}
        \caption{\hspace*{\yval cm} Original}\label{man_a}
    \end{subfigure} %
    \begin{subfigure}{\yval \textwidth}
      \centering
       \includegraphics[width=\xval cm,height=\xval cm]{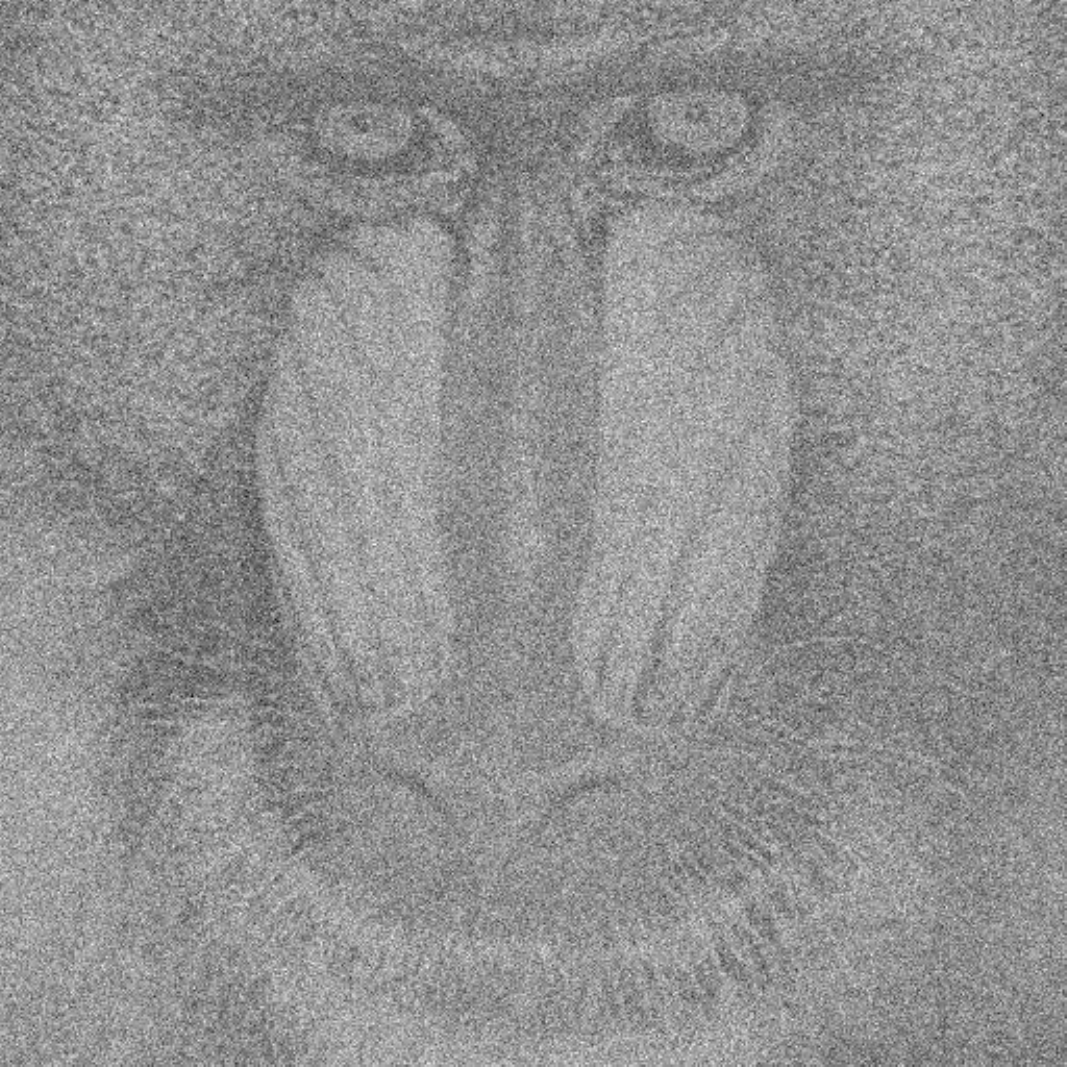}
        \caption{\hspace*{\yval cm} $+$ Gaussian noise}\label{man_b}
    \end{subfigure} %
    \begin{subfigure}{\yval \textwidth}
      \centering
       \includegraphics[width=\xval cm,height=\xval cm]{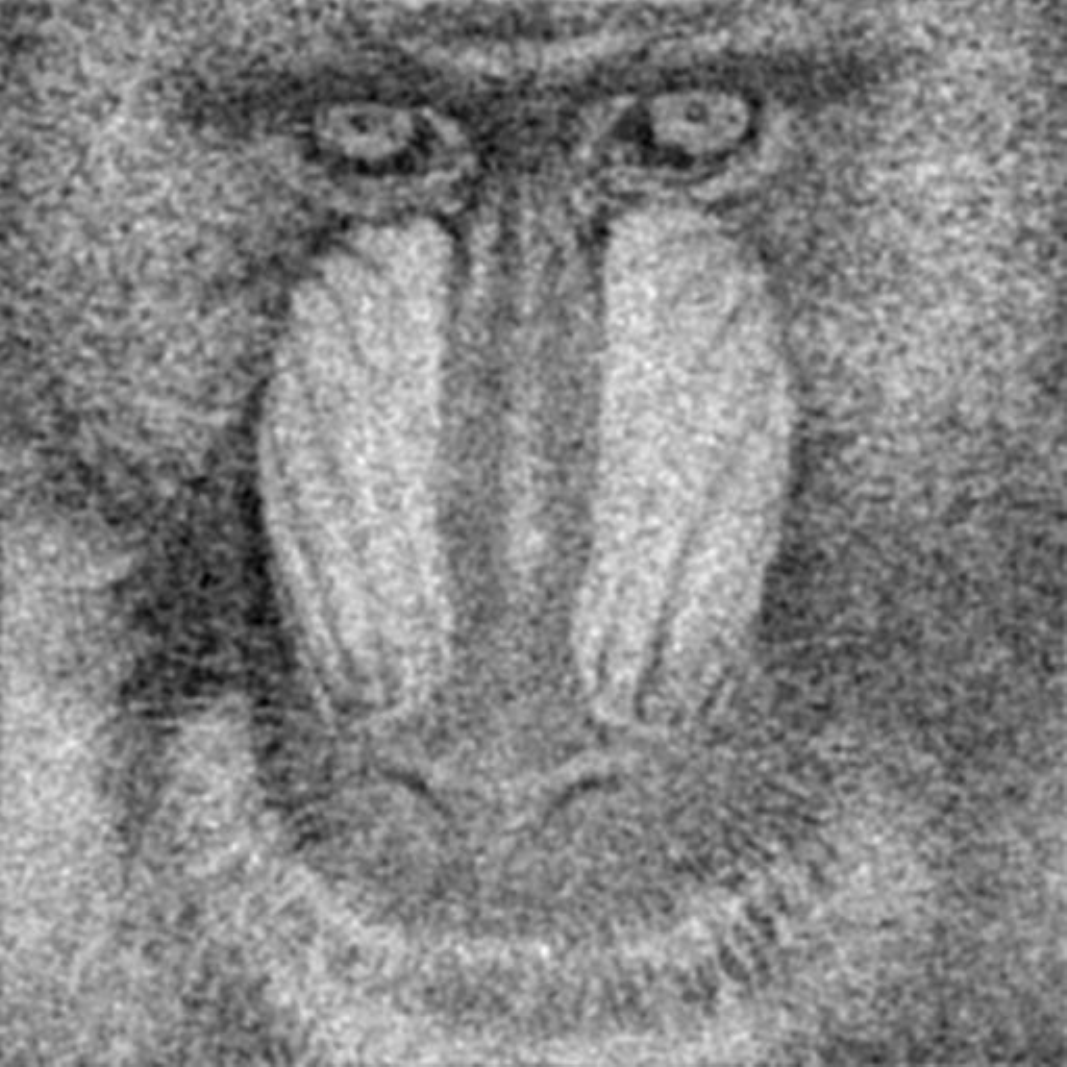}
        \caption{\hspace*{\yval cm} BIFS; $b=1.5$}\label{man_c}
    \end{subfigure} %
    
    \vspace*{0.5cm}
    
    \begin{subfigure}{\yval \textwidth}
      \centering
       \includegraphics[width=\xval cm,height=\xval cm]{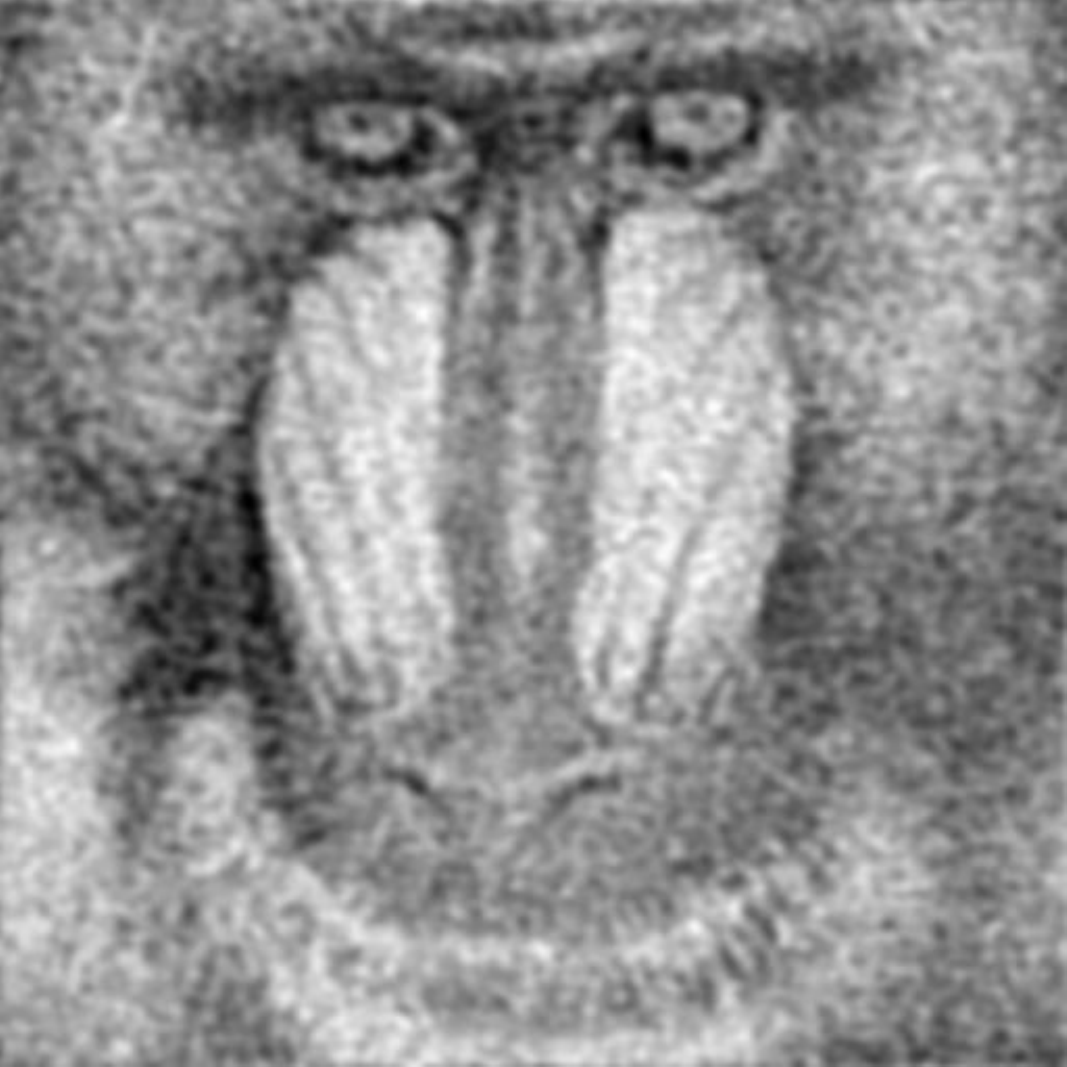}
        \caption{\hspace*{\yval cm} $b=1.75$}\label{man_d}
    \end{subfigure} %
    \begin{subfigure}{\yval \textwidth}
      \centering
       \includegraphics[width=\xval cm,height=\xval cm]{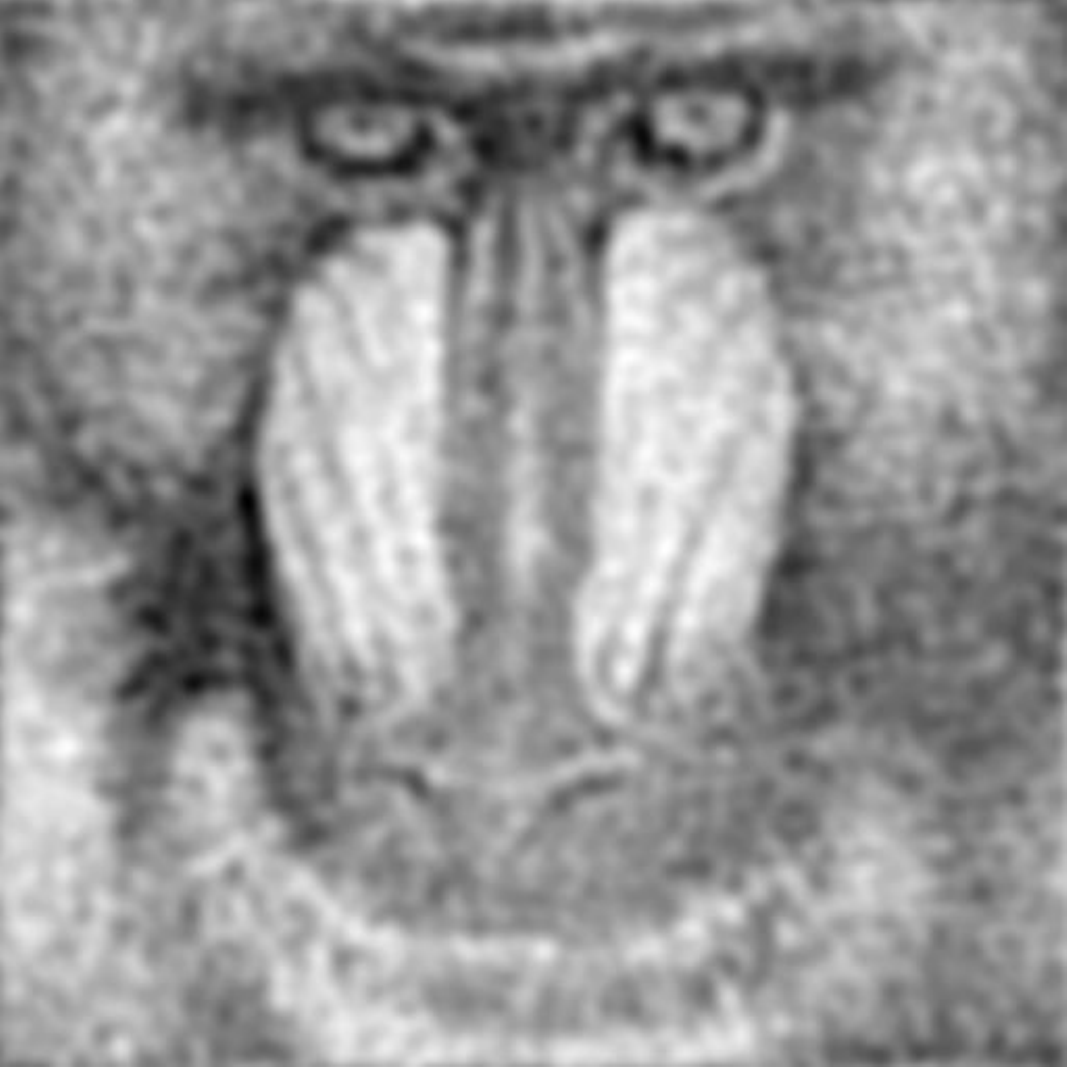}
        \caption{\hspace*{\yval cm} $b=2.0$}\label{man_e}
    \end{subfigure} %
    \begin{subfigure}{\yval \textwidth}
      \centering
       \includegraphics[width=\xval cm,height=\xval cm]{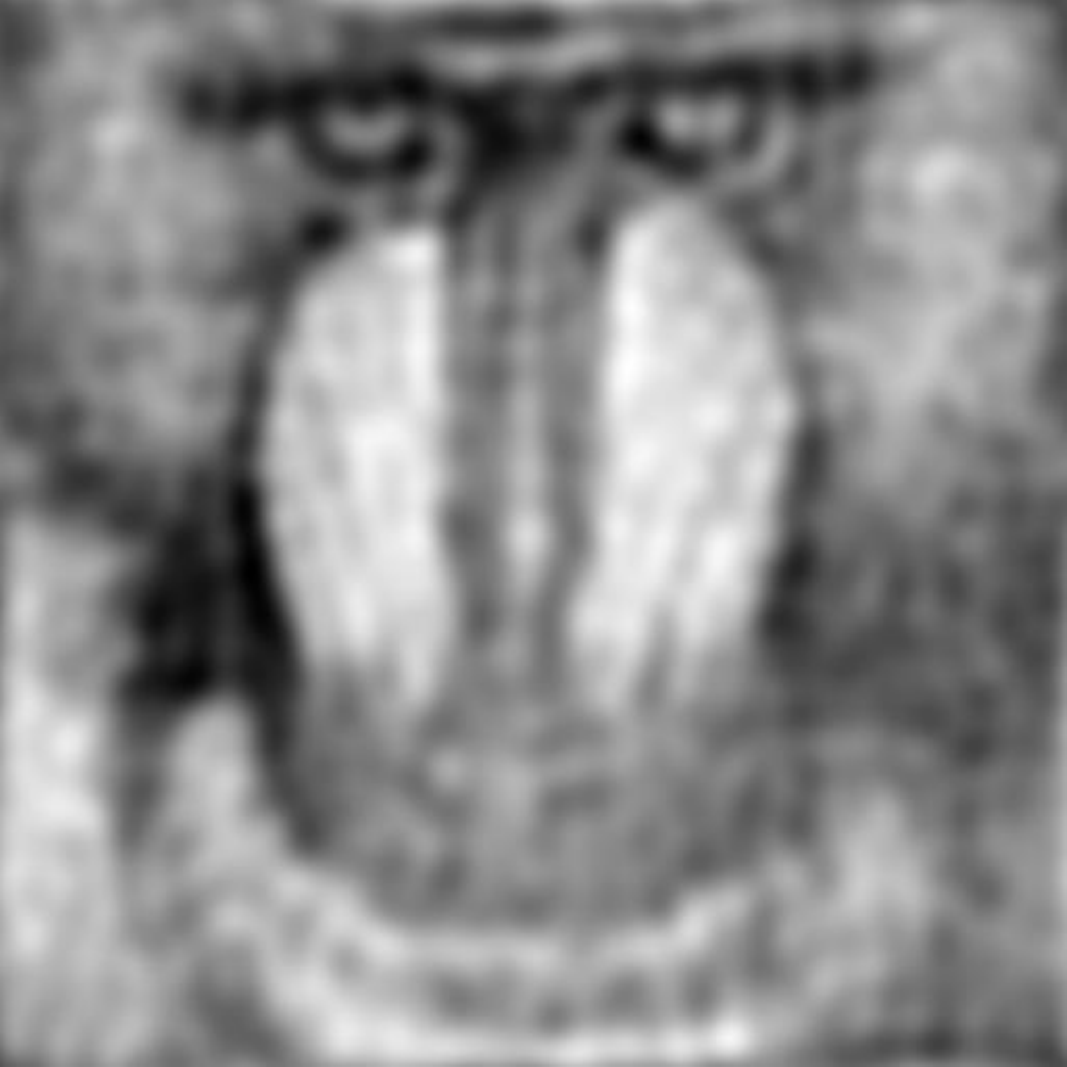}
        \caption{\hspace*{\yval cm} $b=2.5$}\label{man_f}
    \end{subfigure} %
\caption{Mandrill pics, (a) original image; (b) Gaussian noise degraded image; (c) BIFS MAP with $b=1.5$; (d) $b=1.75$; (e) $b=2$; (f) $b=2.5$. \label{mandrillpics} }
\end{figure}

Figure~\ref{mandrillpics} shows the BIFS MAP reconstruction results of a $512 \times 512$ grayscale test image of a Mandrill monkey face using the exponential prior, Rician likelihood, and a parameter function for the exponential distribution mean of the form $f_{\mu}(|k|; a, b) = a/|k|^b$ (inverse exponentiated distance) at all locations except for the origin, $k=(0,0)$, where the prior was an improper uniform distribution on the real line. The top-left panel~(a) shows the original noise-free image and the top-middle panel~(b) shows the same image with added Gaussian noise (zero mean with SD $\approx$ one third of the dynamic range of the original image). The noisy image in panel~(b) is used as the input degraded image into the BIFS models. The remaining panels show BIFS MAP reconstructions for different values of $b$, namely (c)~$b=1.5$, (d)~$b=1.75$, (e)~$b=2$, and (f)~$b=2.5$. The parameter value for $a$ was chosen based on matching the power of the parameter function (sum of square magnitude) to that in the observed data over all Fourier space points other than at $k=(0,0)$. The image intensities in each panel are linearly re-scaled to use the full dynamic range. Re-scaling for comparison is reasonable in situations where the observation of image features is of primary interest, but not where quantification of image intensities is needed.

In examining the MAP reconstructions of Panels~(c)~to~(e) it is clear that the overall level of smoothness increases (and noise suppressed) as the value of $b$ increases. This is to be expected since higher $b$ implies faster decay of the parameter function for the signal modulus with increasing distance from the origin and therefore higher relative intensities at low spatial frequencies. However, the price for higher noise suppression is that finer level features are lost. For example, for $b=2.5$ in Panel~(f) the whiskers and even nostrils of the Mandrill are smoothed away. 

\begin{figure}
\centering
    \begin{subfigure}{\yval \textwidth}
      \centering
       \includegraphics[width=\xval cm,height= \xval cm]{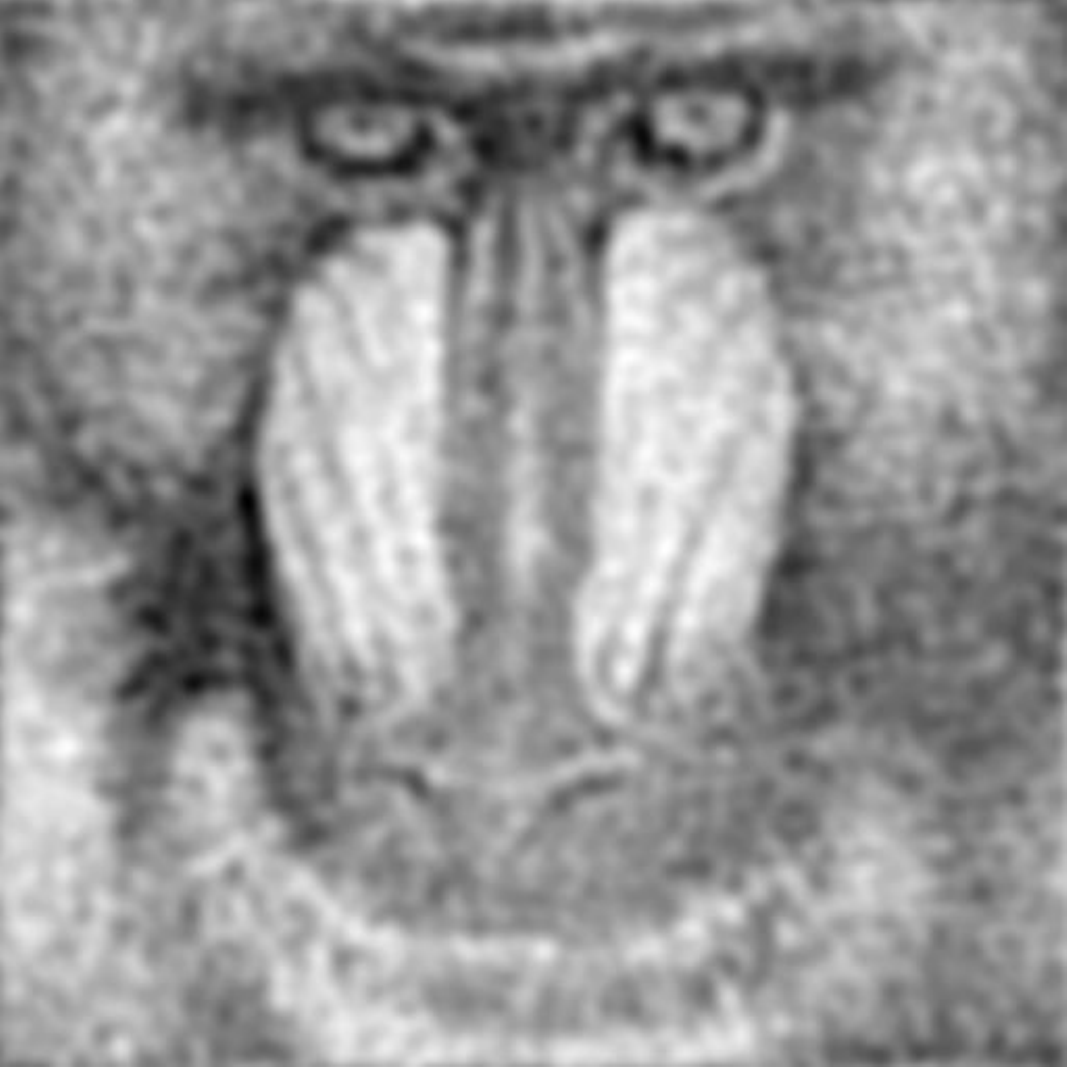}
        \caption{\hspace*{\yval cm} 10 d.f.}\label{tman_10}
    \end{subfigure} %
    \begin{subfigure}{\yval \textwidth}
      \centering
       \includegraphics[width=\xval cm,height=\xval cm]{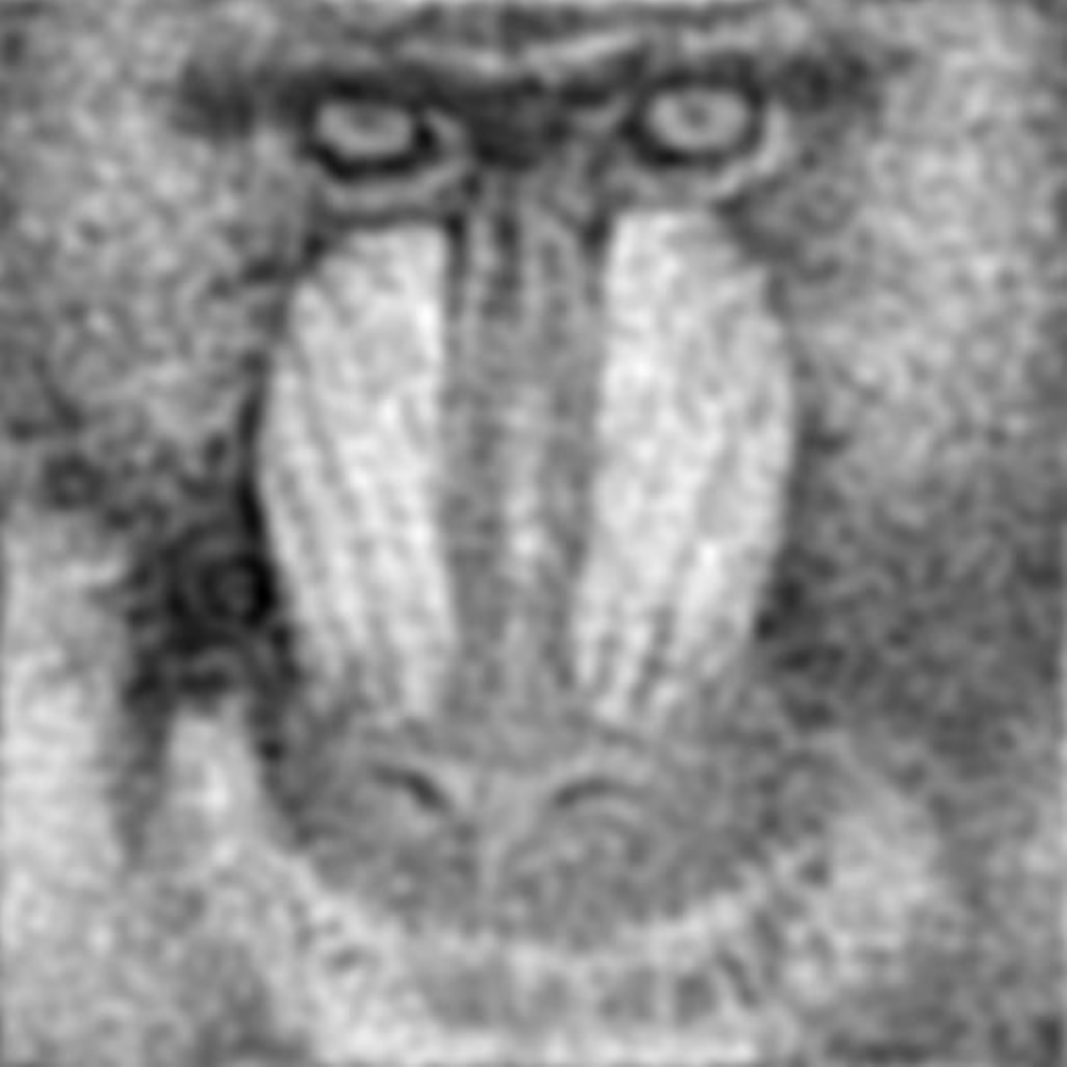}
        \caption{\hspace*{\yval cm} 5 d.f.}\label{tman_5}
    \end{subfigure} %
    \begin{subfigure}{\yval \textwidth}
      \centering
       \includegraphics[width=\xval cm,height=\xval cm]{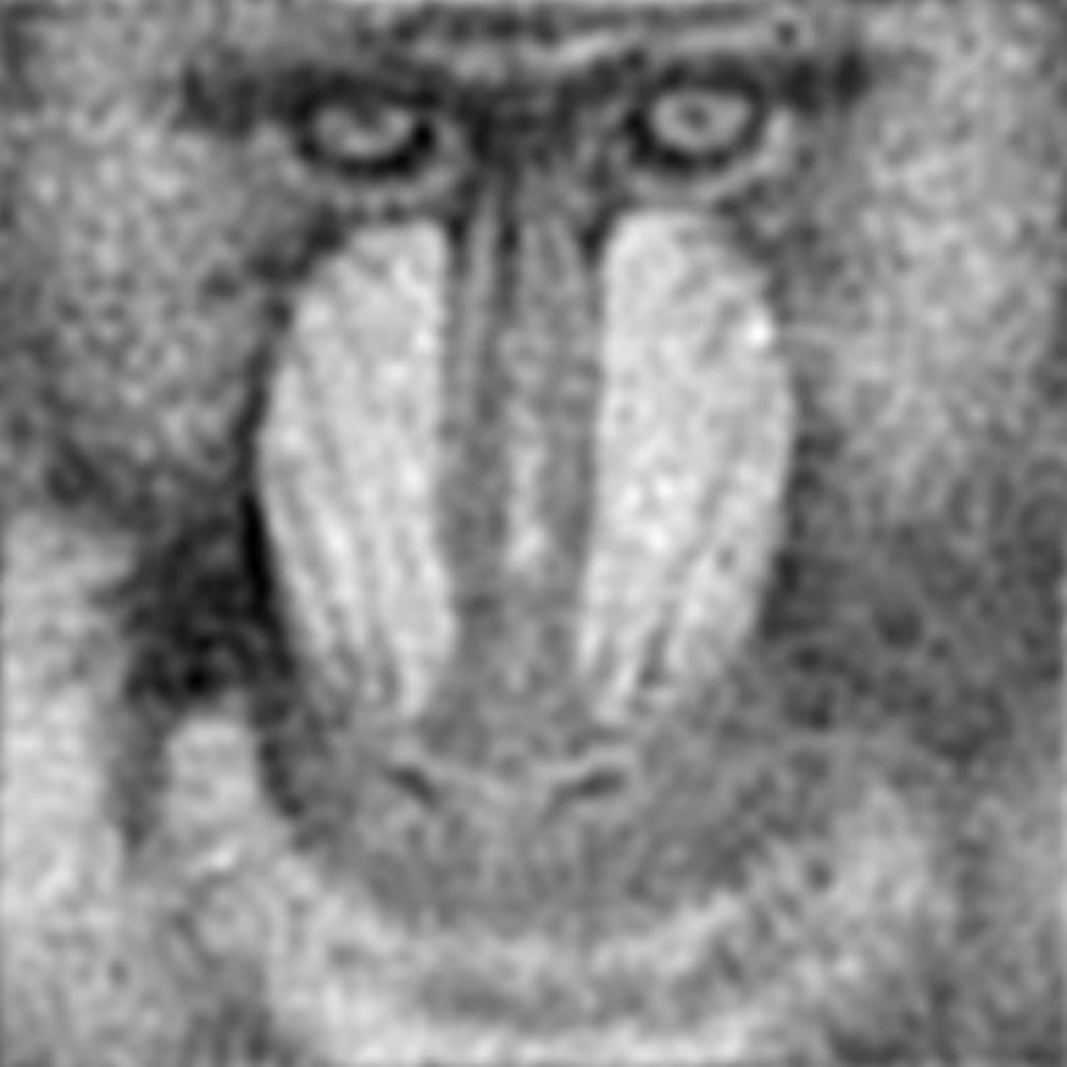}
        \caption{\hspace*{\yval cm} 3 d.f.}\label{tman_3}
    \end{subfigure} %
    
\caption{Mandrill pics reconstructed under mis-specified noise distribution (Student-$t$ mis-specified as normal) all with $b=2$, (a) 10 d.f.; (b) 5 d.f.; (c) 3 d.f. \label{Tmandrillpics} }
\end{figure}

\textbf{Robustness to heavy-tailed noise:} In order to examine whether the BIFS reconstructions were robust to noise distributions with heavier tails, the BIFS procedure was repeated for data with added noise generated from Student $t$-distributions instead of normal, all controlled to have the same SD as the original normal example, but with different degrees-of-freedom (d.f.). Figure \ref{Tmandrillpics} displays the results of the reconstructions using the $b = 2$, but with t-distributed noise with (a)~10~d.f.; (b)~5~d.f.; and (c)~3~d.f. The reconstructions are visually quite similar to each other with slight differences only showing up with the very heavy-tailed 3~d.f. reconstruction. Results from other values of $b$ were also similarly robust.

\subsection{Example 2 -- Frequency band enhancement}

\begin{figure}
\centering
    \begin{subfigure}{\yval \textwidth}
      \centering
       \includegraphics[width=\xval cm,height=\xval cm]{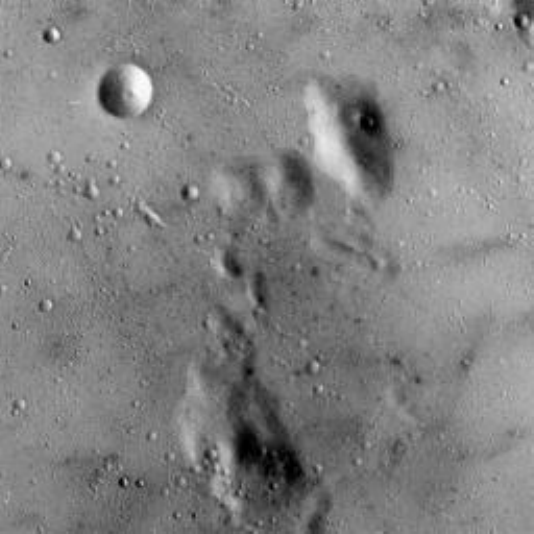}
        \caption{\hspace*{\yval cm} Original}\label{moon_a}
    \end{subfigure} %
    \begin{subfigure}{\yval \textwidth}
      \centering
       \includegraphics[width=\xval cm,height=\xval cm]{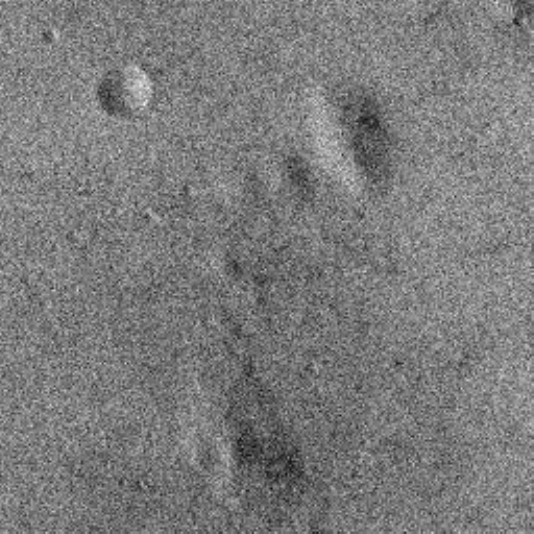}
        \caption{\hspace*{\yval cm} $+$ Gaussian noise}\label{moon_b}
    \end{subfigure} %
    \begin{subfigure}{\yval \textwidth}
      \centering
       \includegraphics[width=\xval cm,height=\xval cm]{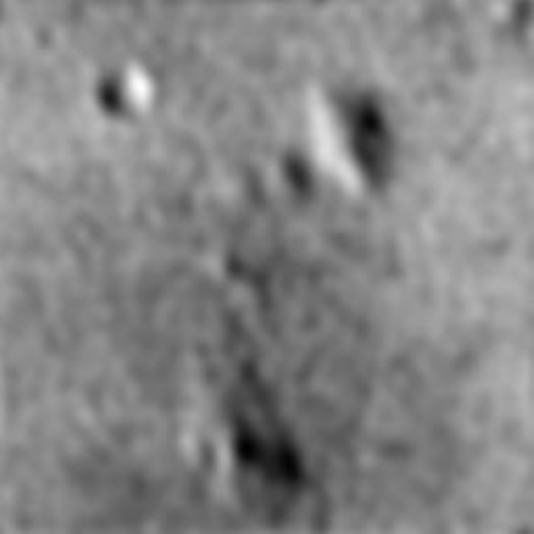}
        \caption{\hspace*{\yval cm} denoising $b=2$}\label{moon_c}
    \end{subfigure} %
    
    \vspace*{0.5cm}
    
    \begin{subfigure}{\yval \textwidth}
      \centering
       \includegraphics[width=\xval cm,height=\xval cm]{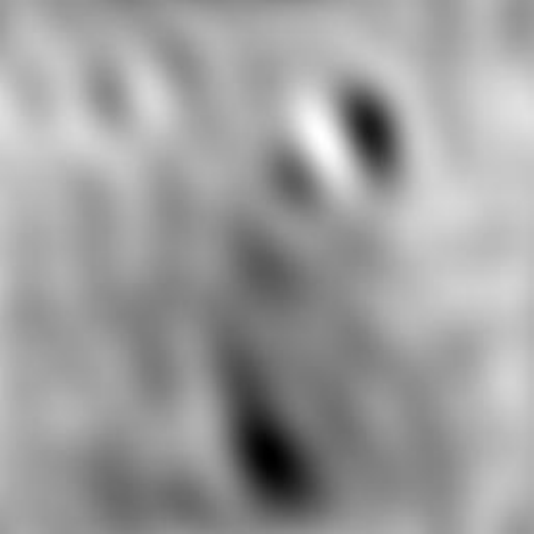}
        \caption{\hspace*{\yval cm} torus $|k| =$ 1 - 5}\label{moon_d}
    \end{subfigure} %
    \begin{subfigure}{\yval \textwidth}
      \centering
       \includegraphics[width= \xval cm,height=\xval cm]{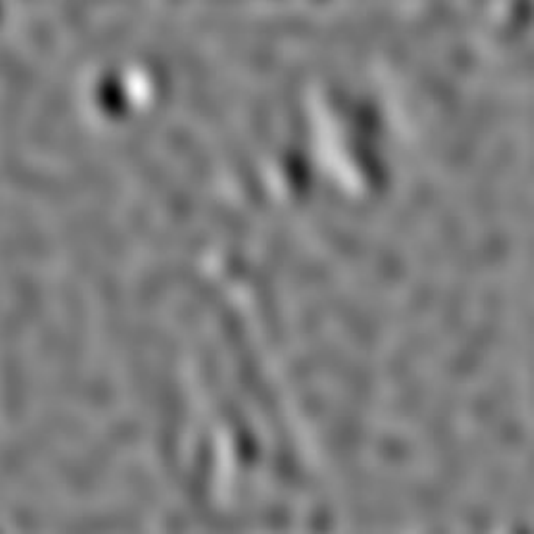}
        \caption{\hspace*{\yval cm} $|k| =$ 10.01 - 15}\label{moon_e}
    \end{subfigure} %
    \begin{subfigure}{\yval \textwidth}
      \centering
       \includegraphics[width=\xval cm,height=\xval cm]{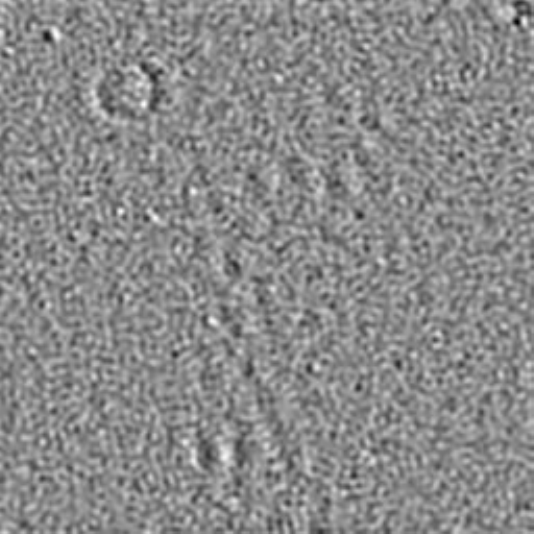}
        \caption{\hspace*{\yval cm} $|k| =$ 15.01 - 60}\label{moon_f}
    \end{subfigure} %
    
    \vspace*{0.5cm}
    
    \begin{subfigure}{\yval \textwidth}
      \centering
       \includegraphics[width=\xval cm,height=\xval cm]{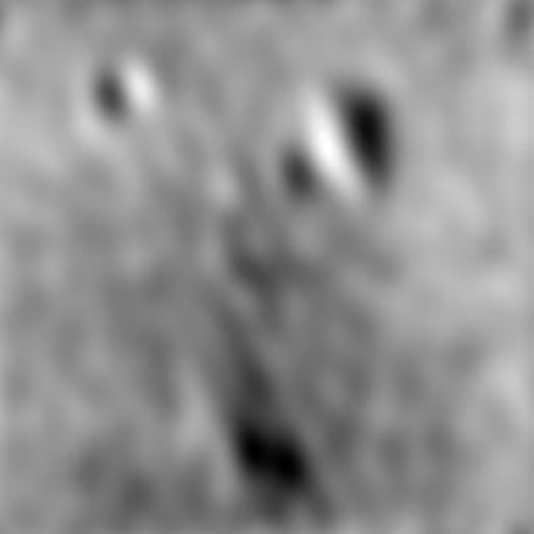}
        \caption{\hspace*{\yval cm} $|k| =$ 1-5 $ \& $ $b=2$}\label{moon_g}
    \end{subfigure} %
    \begin{subfigure}{\yval \textwidth}
      \centering
       \includegraphics[width=\xval cm,height=\xval cm]{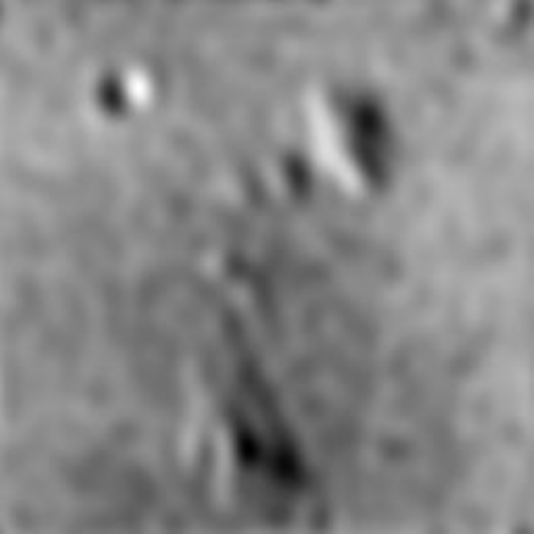}
        \caption{\hspace*{\yval cm} 10.01-15 $ \& $ $b=2$}\label{moon_h}
    \end{subfigure} %
    \begin{subfigure}{\yval \textwidth}
      \centering
       \includegraphics[width=\xval cm,height=\xval cm]{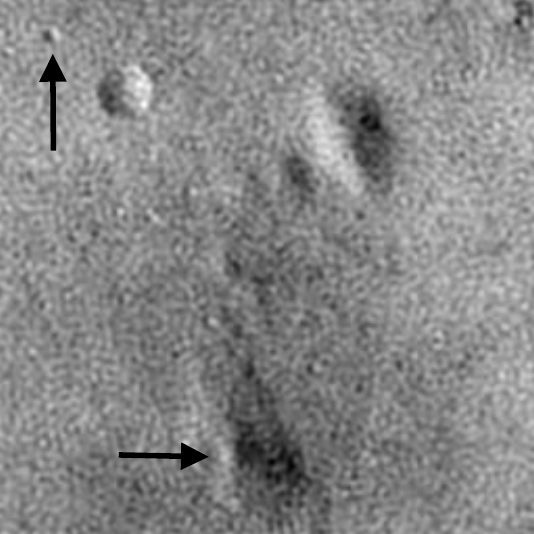}
        \caption{\hspace*{\yval cm} 15.01-60 $ \& $ $b=2$}\label{moon_i}
    \end{subfigure} %
\caption{Moon pics, (a) original image; (b) Gaussian noise degraded image; (c) BIFS MAP estimate using denoising prior from Section~\protect\ref{denoise} with $b=2$; (d) BIFS MAP with "smoothed torus frequencies" based on distance ($|k|$) 1 - 5 from origin; (e) 10.01 - 15; (f) 15.01 - 30; (g) linear combination of smoothed torus 1 - 5 plus denoising prior from Section~\protect\ref{denoise}, i.e., the inverse exponentiated parameter function with $b=2$; (h) 10.01 - 15 plus $b=2$; (i) 15.01 - 60 plus $b=2$. \label{moon1} }
\end{figure}

The example displayed in Figure~\ref{moon1} displays a range of BIFS reconstructions for a $256 \times 256$ grayscale test image of a surface patch on the moon. The reconstructions are based on an exponential prior with Rician likelihood for the parameter function of the signal modulus. Panel~(a) shows the original image and Panel~(b) shows the \iid additive Gaussian noise degraded image serving as the image to apply reconstruction. Panel~(c) displays the BIFS reconstruction based on applying the denoising prior parameter function used in Example~1 with $b=2$. Panels~(d) through~(f) show frequency selective priors for which prior weight is only given to Fourier space locations within a specific range of distances from the origin, i.e., via frequency selective torus parameter functions. These parameter functions are smoothed by an isotropic Gaussian spatial kernel with SD of 1.5 Fourier space pixels in each of the $k_x$ and $k_y$ directions to reduce Gibbs ringing effects caused by sharp cut-offs in frequency selection. For Panel~(d) distances of 1 to 5 pixels from the origin are given non-zero mass; in Panel~(e) 10.01 to 15 pixels; and Panel~(f) 15.01 to 60 pixels. Panels~(g) through (i) show corresponding reconstructions where the parameter function is a weighted average of the torus parameter function directly above and the denoising parameter function of panel~(c); weighted at 90\% torus and 10\% denoising. In all of the examples, the level of each of the parameter functions is adjusted to approximately match total power to that observed in the image data.

Posterior reconstructions using the torus parameter functions are providing results as expected in terms of focusing in on specific frequency windows. The addition of the denoising component in Panels~(g) through~(i) softens the harsh restriction to the specific frequency ranges via a compromise between priors. In particular, for the high range of frequencies of Panel~(f) the addition of the denoising component in Panel~(i) is critical to be able to understand the context of the image. Notice we are able to identify the small crater that is pointed to by the top arrow in Panel~(i) which is missed by the other priors that have less high frequency information. While the features are observable in panel~(f) they are not easily distinguishable as a specific type of object, e.g., compared to the white spot at the middle of the dip in the large crater indicated by the bottom arrow in Panel~(g). 

Note that it would be prohibitively difficult to define neighborhood structures in MRFs to produce priors with the properties of the frequency band enhancement priors. 

There is clearly considerable potential for designing parameter function / prior combinations for BIFS that can produce a range of image processing characteristics that are not readily achievable with MRF priors.

\subsection{Example 3 -- edge enhancement}

\begin{figure}
\centering
    \begin{subfigure}{\yval \textwidth}
      \centering
       \includegraphics[width=\xval cm,height=\xval cm]{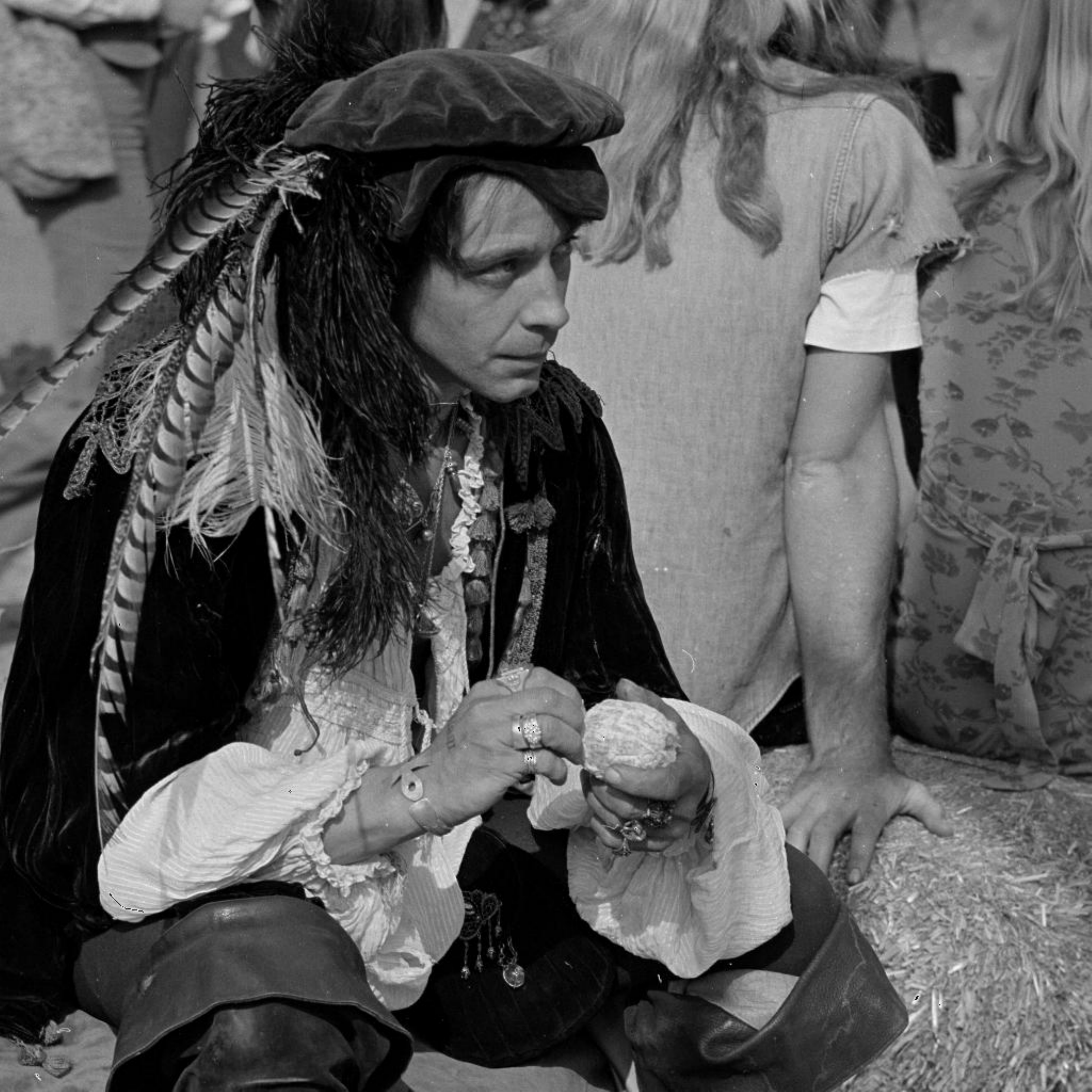}
        \caption{\hspace*{\yval cm} Original}\label{pirate_a}
    \end{subfigure} %
    \begin{subfigure}{\yval \textwidth}
      \centering
       \includegraphics[width=\xval cm,height=\xval cm]{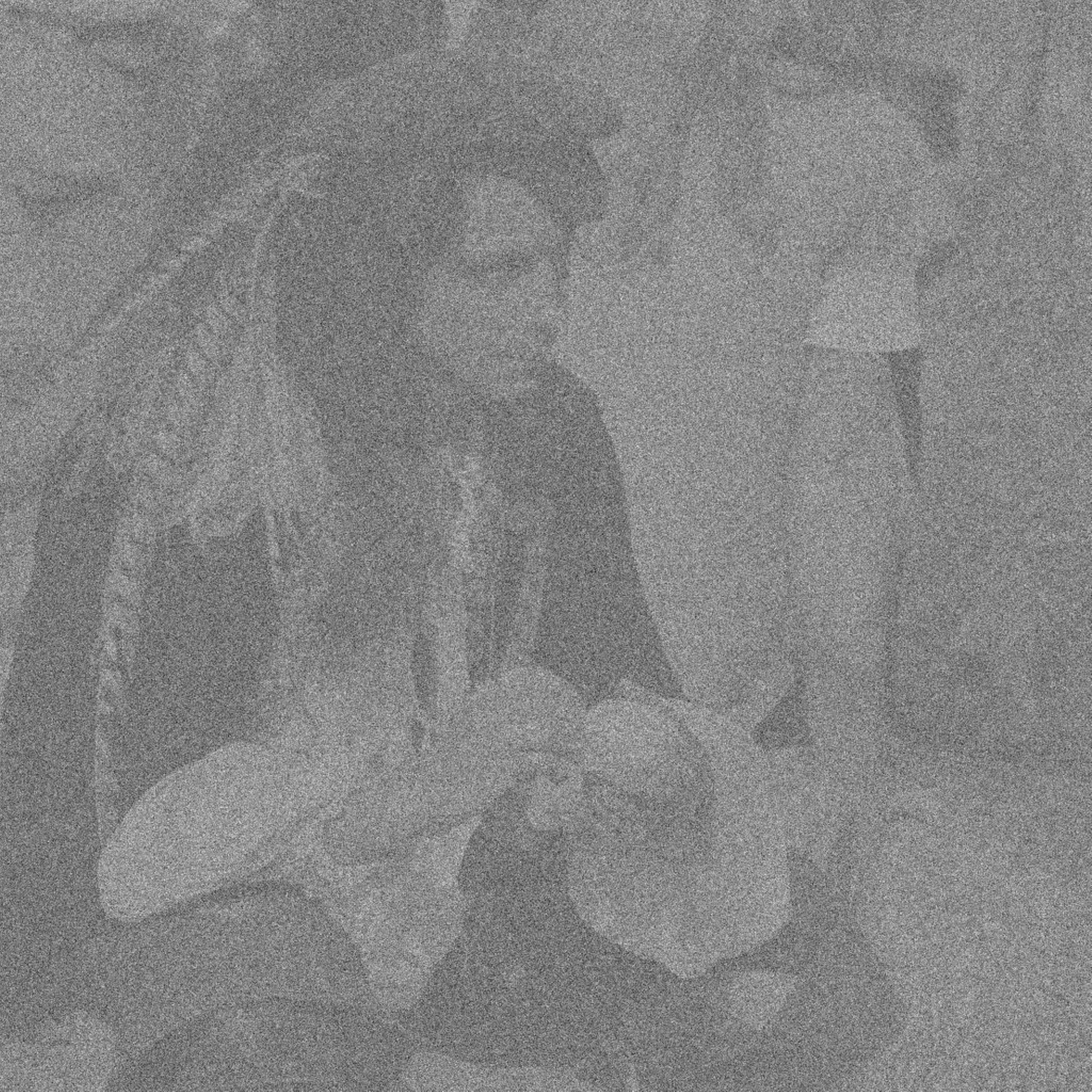}
        \caption{\hspace*{\yval cm} $+$ Gaussian noise}\label{pirate_b}
    \end{subfigure} %
    \begin{subfigure}{\yval \textwidth}
      \centering
       \includegraphics[width=\xval cm,height=\xval cm]{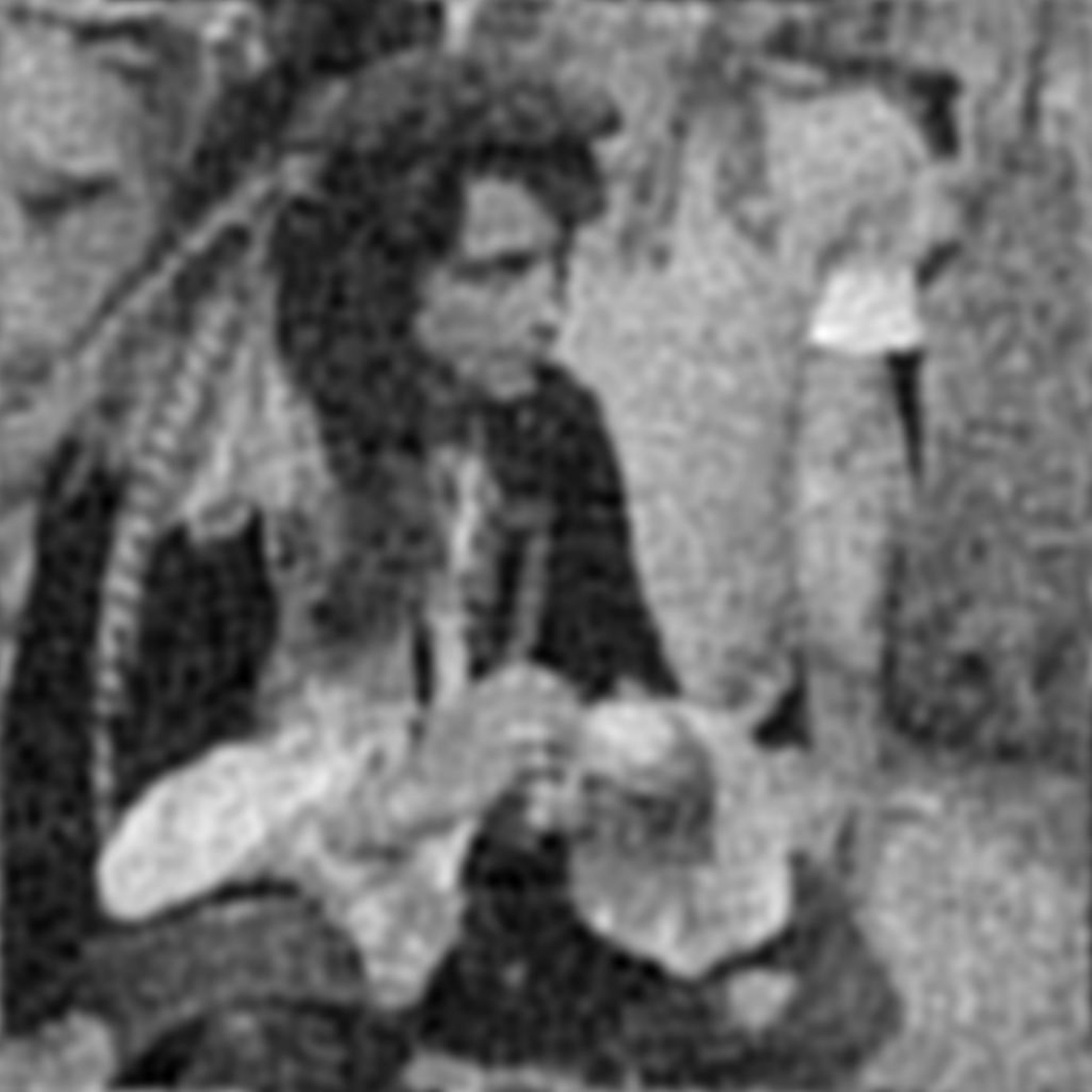}
        \caption{\hspace*{\yval cm} Denoise: $b=2$}\label{pirate_c}
    \end{subfigure} %
    \begin{subfigure}{\yval \textwidth}
      \centering
       \includegraphics[width=\xval cm,height=\xval cm]{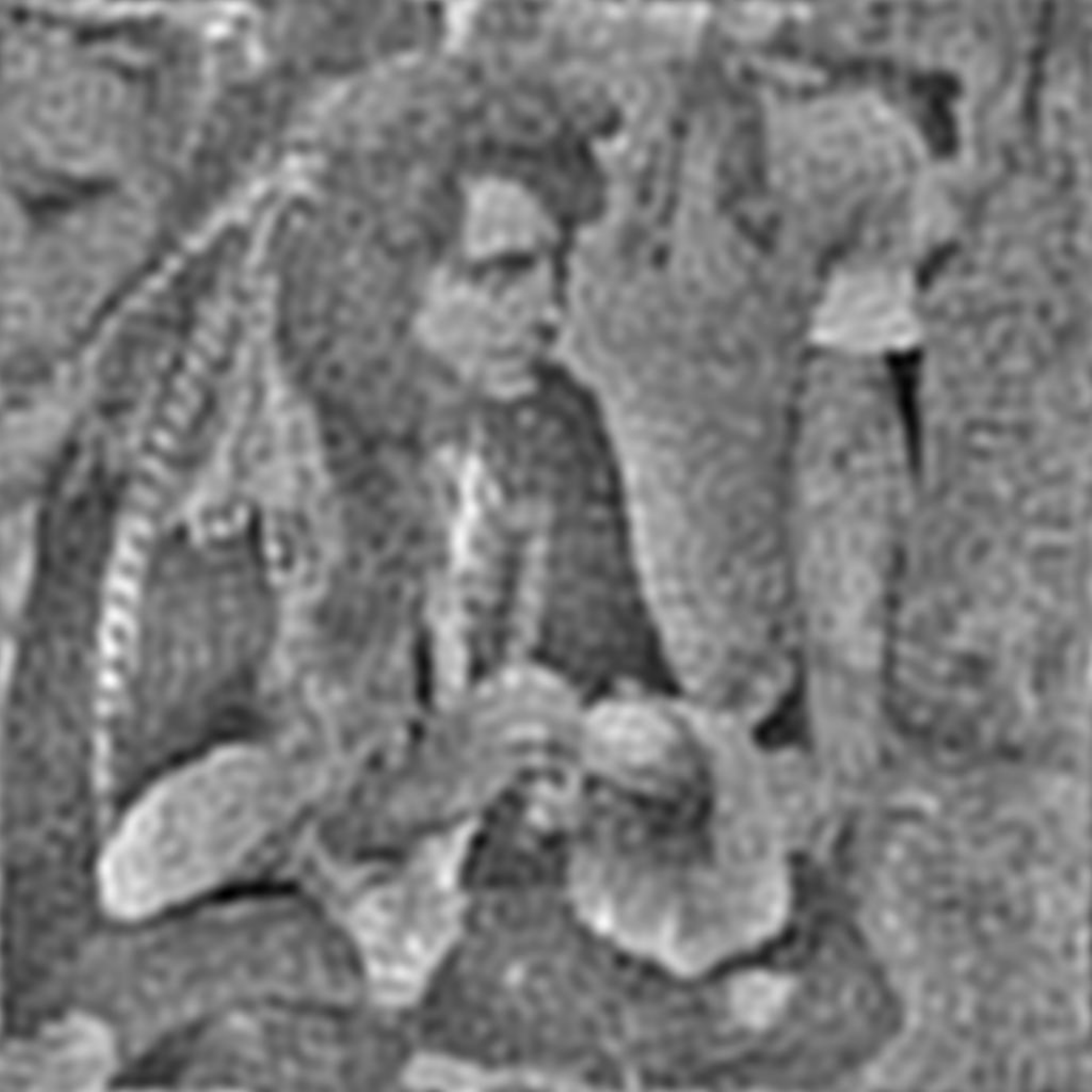}
        \caption{\hspace*{\yval cm} $|k| =$ 10.01 - 50}\label{pirate_d}
    \end{subfigure} %
    \begin{subfigure}{\yval \textwidth}
      \centering
       \includegraphics[width=\xval cm,height=\xval cm]{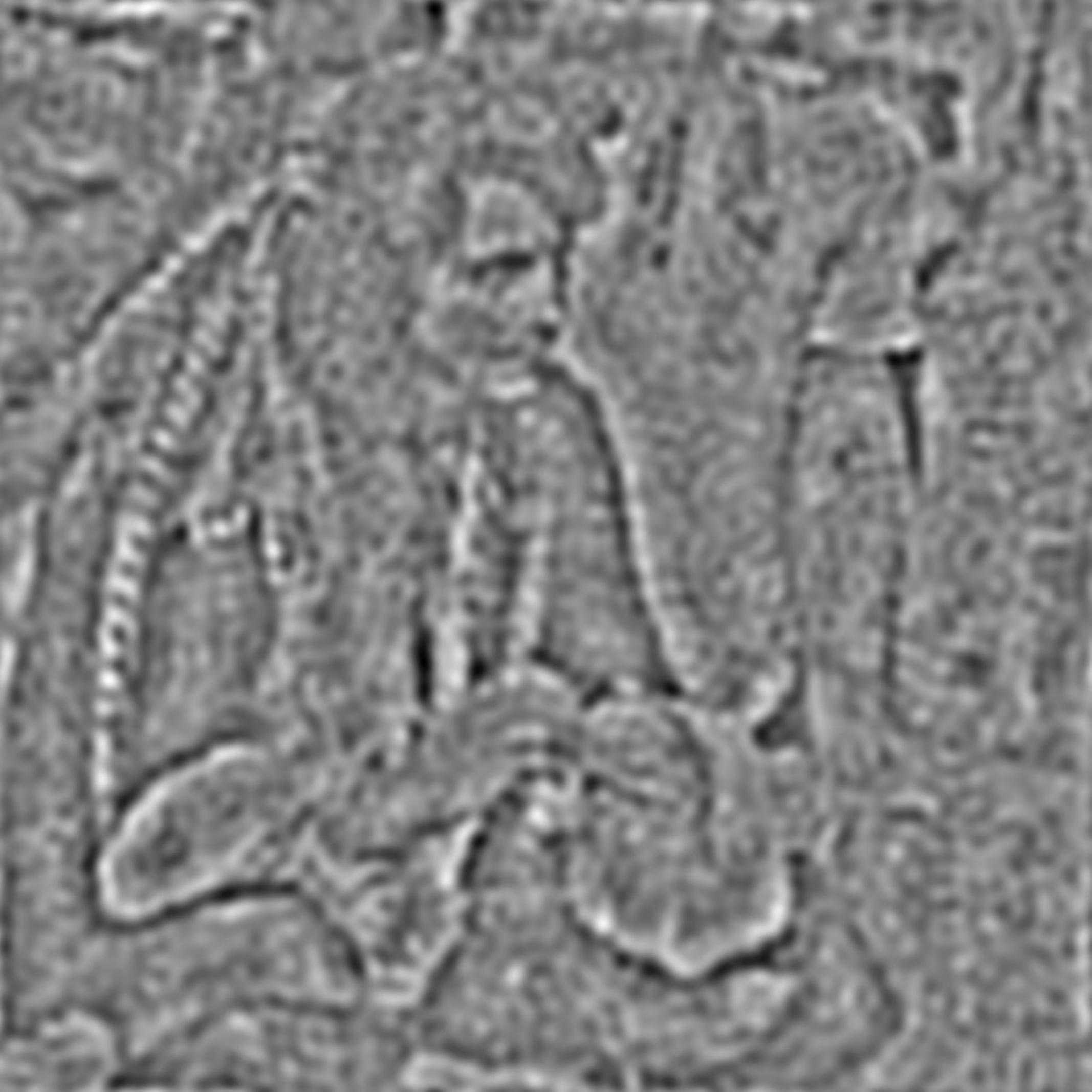}
        \caption{\hspace*{\yval cm} $|k| =$ 15.01 - 50}\label{pirate_e}
    \end{subfigure} %
    \begin{subfigure}{\yval \textwidth}
      \centering
       \includegraphics[width=\xval cm,height=\xval cm]{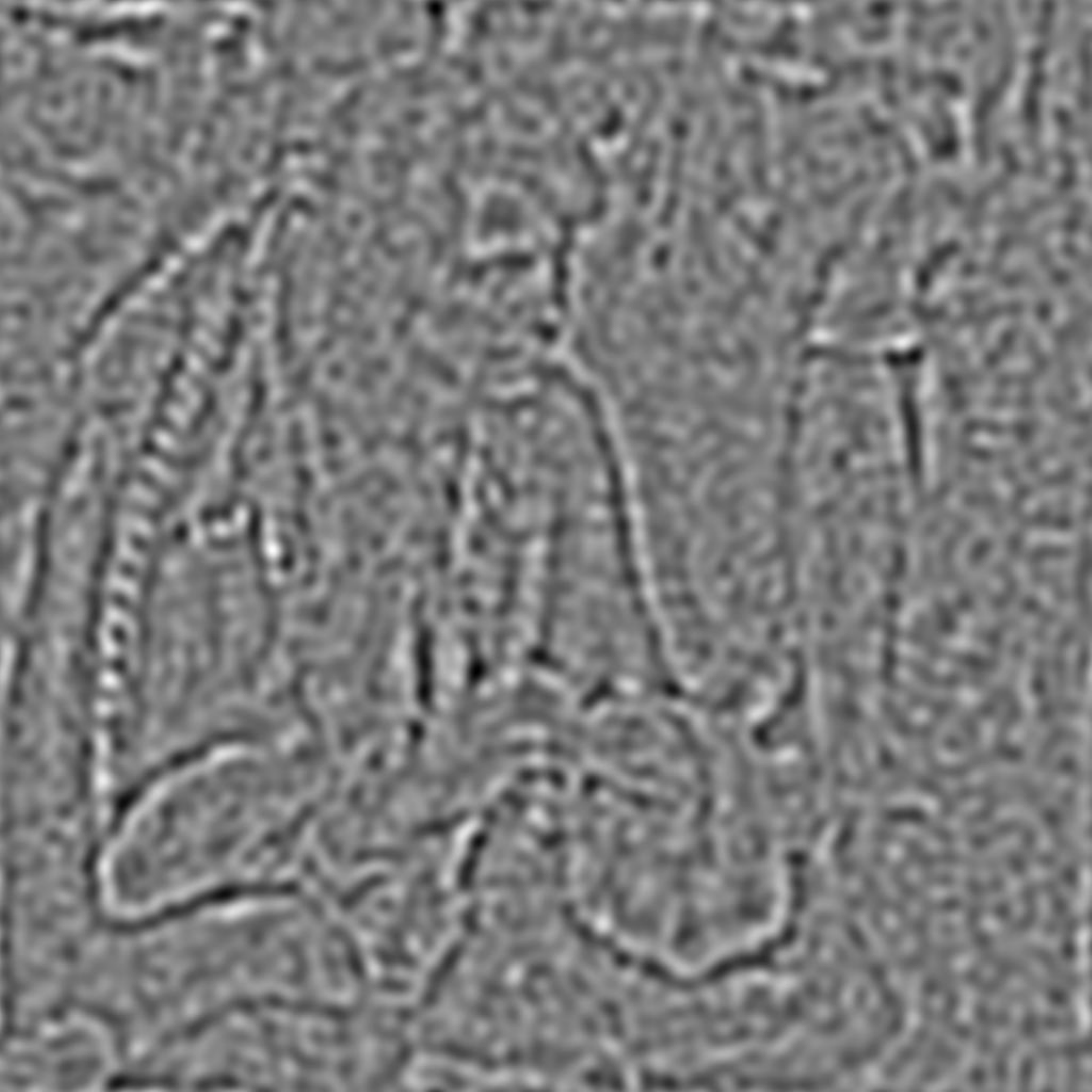}
        \caption{\hspace*{\yval cm} $|k| =$ 20.01 - 50}\label{pirate_f}
    \end{subfigure} %
\begin{subfigure}{\yval \textwidth}
      \centering
       \includegraphics[width=\xval cm,height=\xval cm]{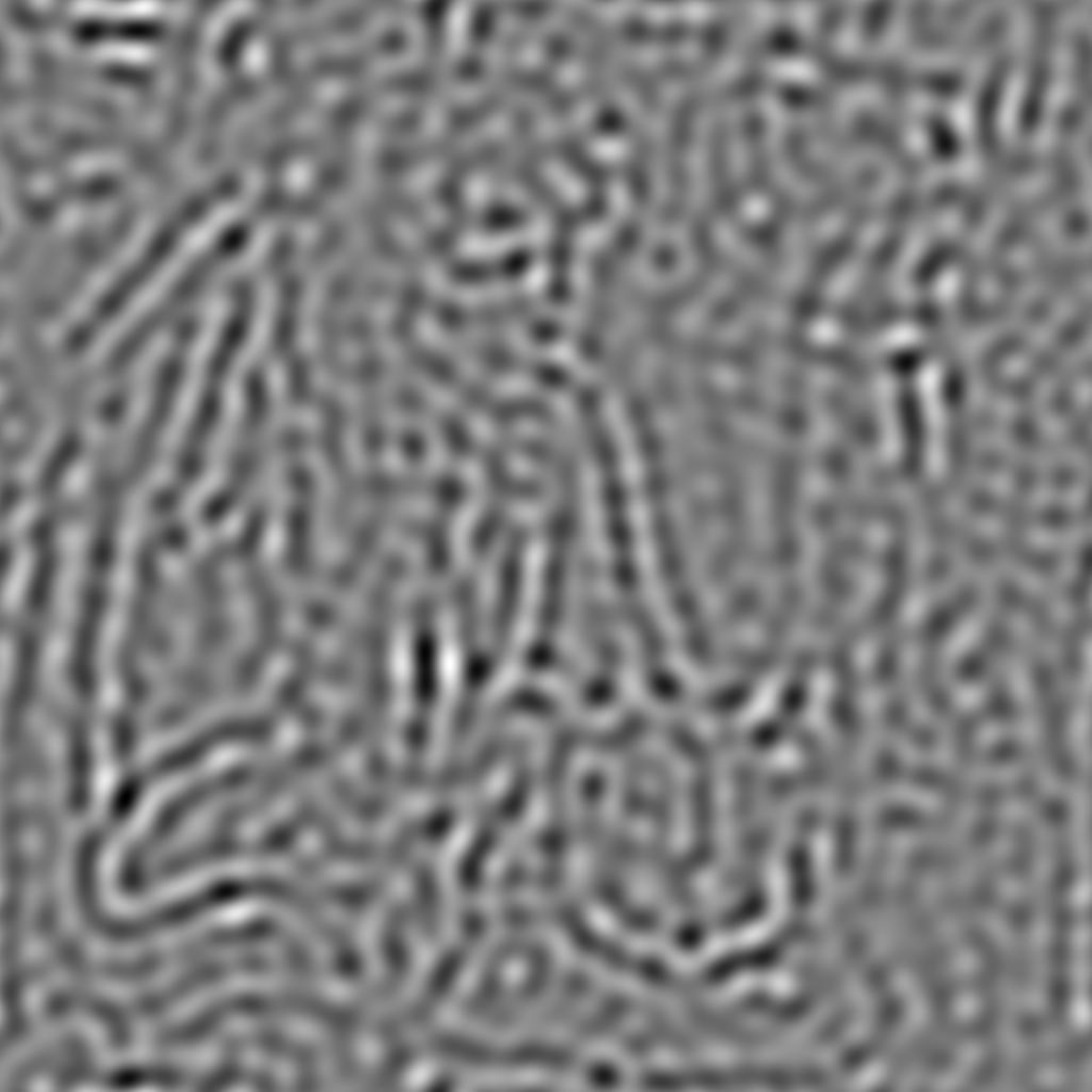}
        \caption{\hspace*{\yval cm} $|k| =$ 20.01 - 30}\label{pirate_g}
    \end{subfigure} %
    \begin{subfigure}{\yval \textwidth}
      \centering
       \includegraphics[width=\xval cm,height=\xval cm]{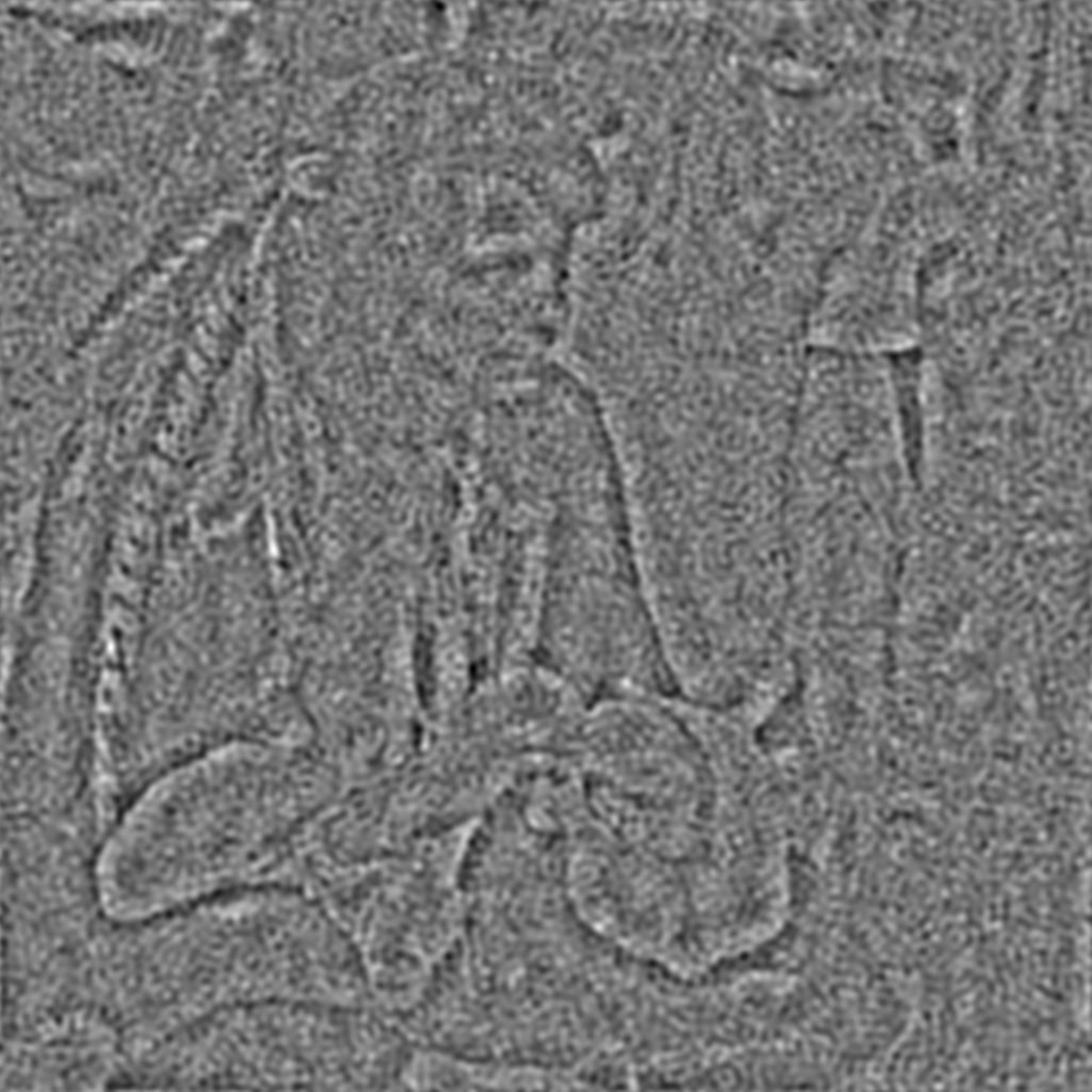}
        \caption{\hspace*{\yval cm} $|k| =$ 20.01 - 100}\label{pirate_h}
    \end{subfigure} %
    \begin{subfigure}{\yval \textwidth}
      \centering
       \includegraphics[width=\xval cm,height=\xval cm]{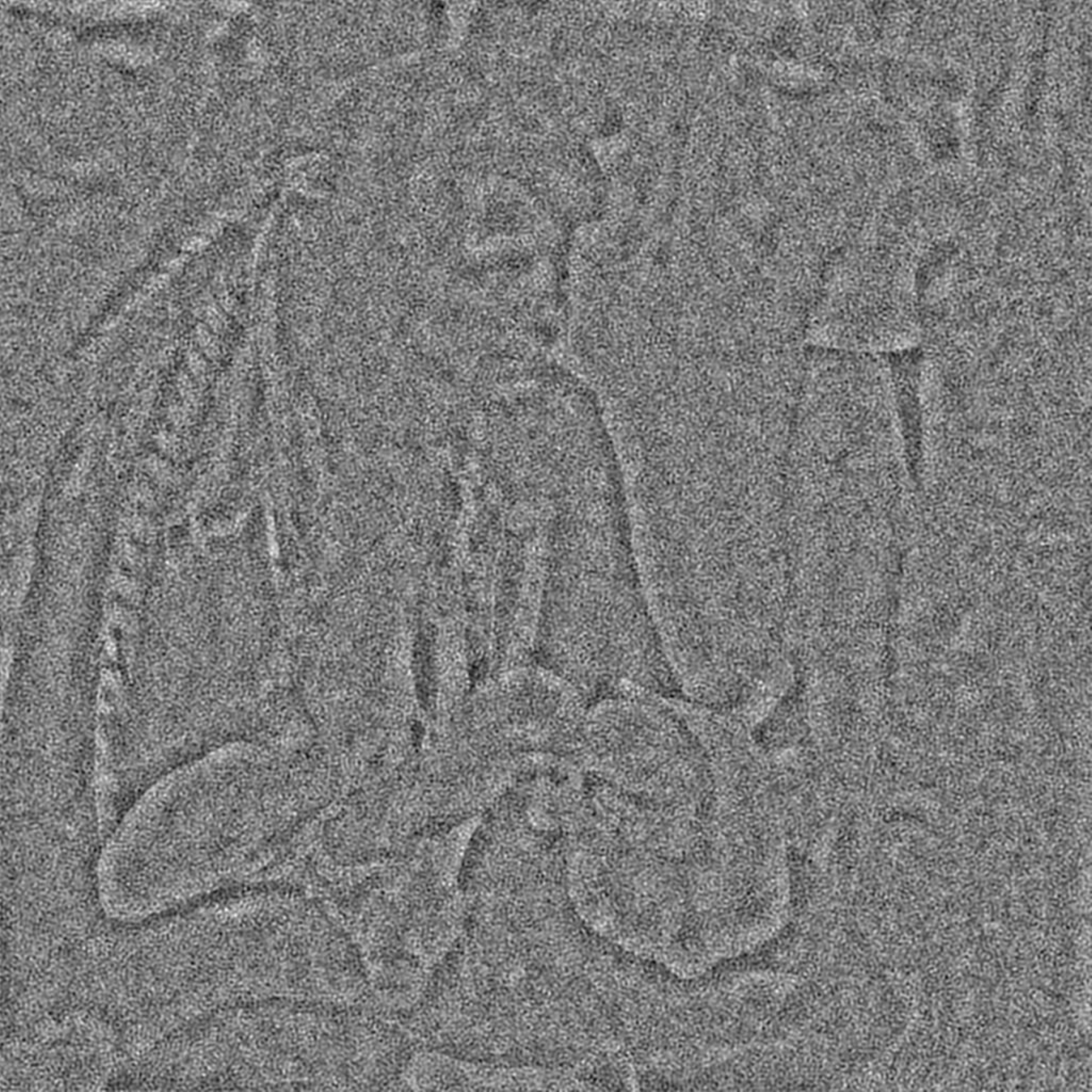}
        \caption{\hspace*{\yval cm} $|k| =$ 20.01 - 150}\label{pirate_i}
    \end{subfigure} %
\caption{Pirate pics, (a) original image; (b) Gaussian noise degraded image; (c) BIFS denoising $b=2$; (d) Frequency enhancing prior with $|k| =$ 1 - 50; (e) 15.01 - 50; (f) 20.01 - 60; (g) 20.01 - 30; (h) 20.01 - 100; (i) 20.01 - 200.  \label{pirate1} }
\end{figure}

The family of frequency enhancing priors described in the previous example can also prove useful for edge enhancement in images as illustrated in this example using a Gaussian noise contaminated version of the standard $1024 \times 1024$ grayscale pirate test image in Figure~\ref{pirate1}. The sequence of images Figure~\ref{pirate1} (d) to (f) shows how the removal of low frequency information isolates edge information. The upper limit on frequencies is chosen as a trade off between capturing the highest frequency edge information vs. potentially masking the edge information if there is too much high frequency noise, as can be seen in Figure~\ref{pirate1} (g) to (i). 

\section{BIFS Properties}  \label{secproperties}

There are multiple properties of the BIFS formulation that prove advantageous relative to conventional MRF-priors and other Bayesian image analysis models:

\subsection{Computational Speed:} \label{compspeed}

The independence specification of the BIFS formulation generally leads to improved computational efficiency over MRF-based or other conventional Bayesian image analysis models that incorporate spatial correlation structures into the priors directly via image space.

For MAP estimation, Bayesian image analysis using MRF or other image space based priors requires high-dimensional iterative optimization algorithms such as iterated conditional modes (ICM) \citep{besag1989digital}, conjugate gradients \citep{hestenes1969multiplier}, or simulated annealing \citep{geman1984stochastic}. Iterative processes are necessary because of the inter-dependence between pixels. In contrast, BIFS simply requires independent posterior mode estimation at each Fourier space location which scales linearly with respect to the number of Fourier space points, i.e. $n$, the number of pixels/voxels. The level of computational complexity is basically $n g(k)$, where $g(k)$ is the complexity of the optimization at each Fourier space point. Therefore, for increased image size or dimension, the computation time is scaled by the proportional increase in the number of Fourier space points, whereas for increased complexity at each Fourier space point to $h(k)$ the total computation is simply scaled by $\frac{h(k)}{g(k)}$.

Computational improvements can similarly be obtained if one wishes to perform posterior mean or other estimation and credible interval generation. MRF and other image-space prior models generally require high-dimensional Markov chain Monte Carlo (MCMC) posterior simulations. However, BIFS only requires low-dimensional MCMC to be performed independently at each Fourier space location. A single realization of the posterior parameter set from each Fourier space location can be inverse Fourier transformed to produce an independent realization of an image from the posterior distribution, thereby avoiding the difficulties of dealing with potentially slow mixing chains. 

Note that for both MAP estimation and posterior sampling, the independence specification of the Fourier space approach also allows for trivial parallelization. Therefore, acceleration of a factor close to the number of processors available can be achieved for up to $n$ processors.  

In addition to direct computational gain, working with BIFS is efficient because it is much easier to try different prior models as defined in Fourier space. In particular, one can very easily change the parameter functions for the prior. Changing the form of the prior distribution at each Fourier space location takes a little more coding effort, but considerably less than writing code for new MRF (or other image-based) posterior reconstructions; trying different MRF priors (beyond simply changing parameter values) is typically a non-trivial task.

Overall, the computational speed afforded by BIFS, in addition to the potential for massive parallelization along with the ease of coding different models, could help widen the use and application of Bayesian image analysis modeling.

\subsection{Resolution invariance} 

Specifying the prior in Fourier space allows straightforward adjustment to the prior when image resolution is changed to retain the same properties over the field of view for frequencies that can be specified at both resolutions. When increasing resolution in image space by factors of two, the central region of the Fourier transform at higher resolution corresponds very closely to that of the complete Fourier transform at lower resolution; because they correspond to the same spatial frequencies within the field of view. The lower resolution image is very close to a band-limited version of the image at higher resolution. Note that this is only approximate as the increased resolution can slightly alter the magnitude of the measured Fourier components whose frequency is significantly lower than that of the resolution. However, this effect can be bounded by the ratio of the maximum resolution to that of the wavelength of the measured frequency so is typically quite small.  This small change in magnitude will remain small in the posterior estimate when the parameter function is continuous.

When the resolution increases but not by a factor of two, the same approach of matching  parameter functions over the frequency range can still be applied. However, there will no longer be a direct match of points over the lower frequencies and therefore there will be a less perfect match of the prior distributions overall. 

The above described BIFS approach to matching over different resolutions contrasts with MRF models, for which in order to retain the spatial properties of the prior at lower frequencies, an increase in resolution would require careful manipulation of neighborhood structure and prior parameters to match spatial auto-covariance structures between the different resolution images~\citep{rue2002fitting}.

\subsection{Isotropy} \label{isotropy}

In order to specify an approximately isotropic BIFS prior, all that is required is for the prior to be specified in such a way that the distribution at each Fourier space location only depends on the distance from the center of Fourier space. This can be achieved by defining the parameter functions for the prior completely in terms of distance from the origin in Fourier space, i.e., $\pi(\ft x_k)  = g(|k|)$, and not the orientation with respect to the center of Fourier space. (The "approximately" qualifier is needed because the prior will be isotropic up to the maximal level afforded by the discrete -- and anisotropic -- specification of Fourier space on a regular, i.e. square, grid.) The relative ease with which isotropy is specified can be contrasted with that of MRF-based priors where local neighborhood characteristics need to be carefully manipulated by increasing neighborhood size and adjusting parameter values to induce approximate spatial isotropy~\citep{rue2002fitting}. Note however, that even for BIFS priors it may still be possible to adjust the parameter function to lead to greater isotropy in practice by tweaking the parameter function to undo anisotropy effects due to Fourier space discretization, though it is not obvious how one might go about achieving this.

 Although easy to specify as such, isotropy is clearly not a requirement of the BIFS prior formulation. Anisotropy can be induced by allowing the parameter functions for the prior to vary  along different orientations.

\section{Approximating Markov random fields with BIFS} \label{secapprox}

Given that MRF priors have become something of a standard in Bayesian image analysis, it would be useful to generate BIFS models to try and match these commonly used priors. We propose an approach to specifying a prior that is close to an MRF of interest.

To find a formulation in Fourier space that approximately matches a corresponding MRF model we propose taking the steps described in Algorithm \ref{simMRFalg}. 
\begin{algorithm}
\caption{Simulating MRF models with BIFS} \label{simMRFalg}
\begin{algorithmic}[1]
\State Simulate a set of images from the MRF prior distribution and take the FFT of each image
\State Determine a model for the prior probability distribution of the modulus to be used and independently estimate its' parameters at each Fourier space location
\State Determine an appropriate parameter function form over Fourier space for each of the parameters from the chosen prior based on estimates from the simulated data. (If maximally isotropic approximations to the prior are required then the parameter functions need to be chosen subject to the constraint that it only depends only on distance from the origin of Fourier space)
\State Estimate the coefficients of the parameter function by fitting to the marginally estimated parameters in Fourier space, e.g. via least squares
\end{algorithmic}
\end{algorithm}

Note that it is possible to take this approach when estimating a BIFS approximation to any Bayesian image analysis prior model, though the potential to match higher-level priors (e.g. ones that directly model geometric properties of objects in the image) is likely to be limited. 

\subsection{Example 4 - Matching Gaussian MRFs} \label{MRImatch}

 We approximate the simple first-order pairwise difference intrinsic Gaussian MRF (IG-MRF) 
 \begin{equation*}
     \pi(x) \propto \exp \left\{ - \frac{\kappa}{2} \sum_{i \sim j} \left(x_i-x_j \right)^2 \right\}
 \end{equation*} 
 as described in \cite{besag1989digital}. The sum over $i \sim j$ is over all unordered pairs of pixels such that $i$ and $j$ are vertically or horizontally adjacent neighbors. The model is called \textit{intrinsic} because the overall mean is not defined and therefore the prior is non-stationary and improper with respect to the overall mean level. 
 
 We used the R-INLA package \citep{rue2017bayesian} to simulate 1,000 realizations of a $128 \times 128$ first-order IG-MRF with $\kappa = 1.0$ and first-order neighborhood structure wrapped on a torus. At each Fourier space location we adopted a prior distribution for the modulus such that the square of the modulus is distributed as exponential. This choice of prior falls in line with the theoretical results in \citep{rue2005gaussian}, Ch~2.6 for the specific case of simulating a Gaussian Markov random field with neighborhood structure wrapped on a torus (i.e. with block-circulant precision matrix); details and posterior estimate derivation is given in the online supplementary Appendix.

\begin{figure}
    \centering
    \includegraphics[width=8 cm,height=5 cm]{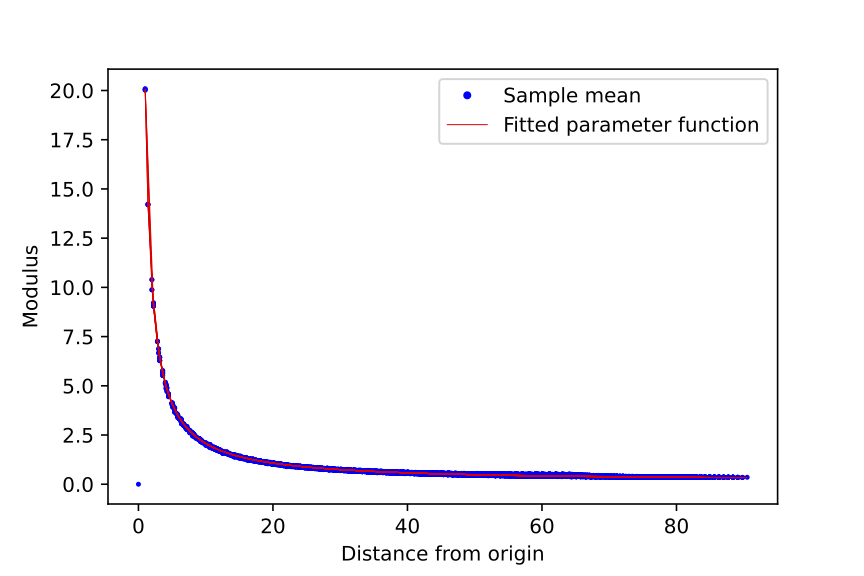}
    \caption{Parameter model fit for modulus as a function of distance from the origin of Fourier space}
    \label{fig:GMRFmodFit}
\end{figure}

Instead of exactly matching the MRF using the eigenvalues to model the IG-MRF in Fourier space as in \citep{rue2005gaussian}, we instead fit an approximately isotropic parameter function to the mean of the modulus at each Fourier space location using least squares. The form of the parameter function used was $$f(|k|; \textbf{a}) = \frac{a_0}{a_1 + a_2 |k| + a_3 |k|^2 + a_4|k|^3}$$ (chosen based on a trial-and-error approach to get a good least squares fit as in Figure~\ref{fig:GMRFmodFit}).  

\begin{figure}
    \begin{subfigure}{\yval \textwidth}
      \centering
       \includegraphics[width=\xval cm,height=\xval cm]{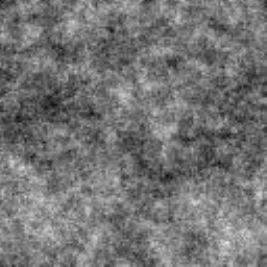}
        \caption{\hspace*{\yval cm} GMRF sim}\label{gsim_a}
    \end{subfigure} %
    \begin{subfigure}{\yval \textwidth}
      \centering
       \includegraphics[width=\xval cm,height=\xval cm]{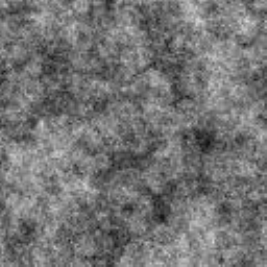}
        \caption{\hspace*{\yval cm} BIFS sim}\label{gsim_b}
    \end{subfigure} %
    \begin{subfigure}{\yval \textwidth}
      \centering
       \includegraphics[width= 4.5 cm,height= 3.5 cm]{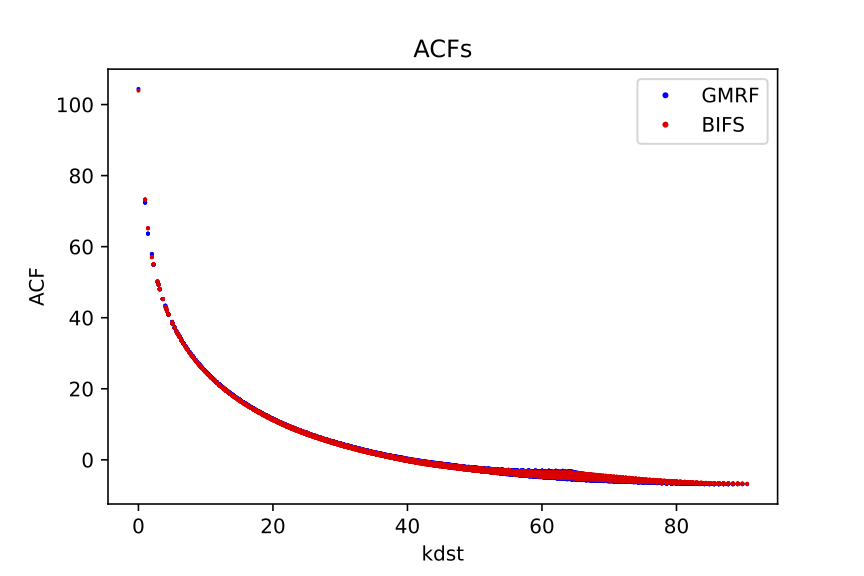}
        \caption{\hspace*{\yval cm} ACFs}\label{acf_c}
    \end{subfigure} 
    \caption{Example simulations for the first-order intrinsic GMRF using a direct approach in R-INLA in panel~(a) and the BIFS approximation with isotropic parameter function in panel~(b). The estimated ACF as a function of distance for each of the simulated models is given in panel~(c).}
    \label{simMatch}
\end{figure}

Simulations from the BIFS prior were well-matched to their MRF counterpart. Figure~\ref{simMatch} shows example realizations from each of the BIFS and direct MRF simulations in panels~(a) and~(b) respectively and they clearly exhibit similar properties.  Panel~(c) shows the respective estimated autocovariance functions (ACFs) as a function of distance based on 1,000 simulations of each random field. The spread observed in the estimated ACFs at longer distances is due to the anisotropy of the processes induced by the rectangular lattice and for the MRF because of the anisotropic representation of the neighborhood structure; hence the slightly narrower band for the BIFS simulations (see Section~\ref{isotropy}).  

\begin{figure}
\centering
    \begin{subfigure}{\yvalb \textwidth}
      \centering
       \includegraphics[width=\xval cm,height=\xval cm]{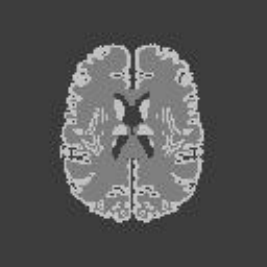}
        \caption{\hspace*{\yval cm} True signal}\label{gmrf_a}
    \end{subfigure} %
    \begin{subfigure}{\yvalb \textwidth}
      \centering
       \includegraphics[width=\xval cm,height=\xval cm]{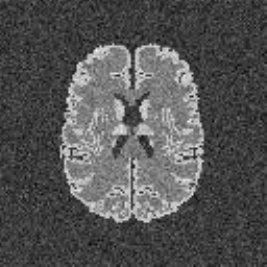}
        \caption{\hspace*{\yval cm} $+$ noise}\label{gmrf_b}
    \end{subfigure} %
    \begin{subfigure}{\yvalb \textwidth}
      \centering
       \includegraphics[width=\xval cm,height=\xval cm]{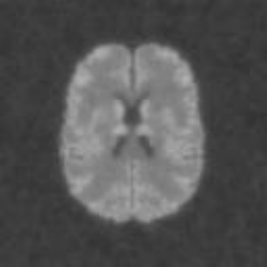}
        \caption{\hspace*{\yval cm} IG-MRF MAP}\label{gmrf_c}
    \end{subfigure} %
    \begin{subfigure}{\yvalb \textwidth}
      \centering
       \includegraphics[width=\xval cm,height=\xval cm]{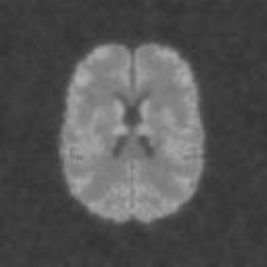}
        \caption{\hspace*{\yval cm} BIFS MAP}\label{gmrf_d}
    \end{subfigure} %
    \begin{subfigure}{\yvalb \textwidth}
      \centering
       \includegraphics[width=\xval cm,height=\xval cm]{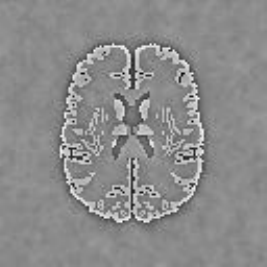}
        \caption{\hspace*{\yval cm} IG-MRF residuals}\label{gmrf_e}
    \end{subfigure} %
    \begin{subfigure}{\yvalb \textwidth}
      \centering
       \includegraphics[width=\xval cm,height=\xval cm]{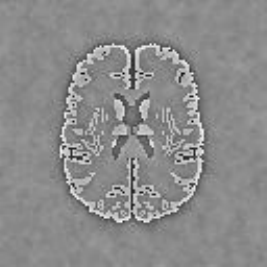}
        \caption{\hspace*{\yval cm} BIFS residuals}\label{gmrf_f}
    \end{subfigure} %
\caption{BIFS match to IG-MRF prior for perfusion MRI simulation study: (a) simulated brain signal image; (b) with added Gaussian noise; (c) IG-MRF MAP reconstruction; (d) BIFS MAP approximation; (e) IG-MRF MAP residuals; (f) BIFS MAP residuals. Panels~(a) to~(d) are normalized to be on the same dynamic scale. Panels~(e) and~(f) are matched to a dynamic scale over the range of the residuals in both images.}
\label{MRImatchFig}
\end{figure}

To compare posterior estimates between the two approaches we examine a $128 \times 128$ simulated image generated using a version of the Montreal Neurological Institute (MNI) brain \citep{evans19933d} that had been segmented into gray matter (GM), white matter (WM), and cerebro-spinal fluid (CSF). GM is the outer ribbon (the cortex) around the brain where neuronal activity occurs, WM is made up of the connective strands that enable different regions of the cortex to communicate with each other, and CSF is fluid in the brain. The signal was generated with intensity 20.0 in GM, 10.0 in WM and 0.0 elsewhere and is displayed in Panel~(a) of Figure~\ref{MRImatchFig}. Gaussian noise (\iid) with SD of 2.5 was added to generate a degraded image as displayed in Panel~(b).

The MAP estimate for the reconstructed image was obtained using each of the IG-MRF prior (using conjugate gradients optimization) and the BIFS approximation. The IG-MRF and BIFS reconstructions (Panels~(c) and~(d), respectively) are indistinguishable visually. This is confirmed when looking at the residual maps in panels~(e) and~(f) which are also indistinguishable.  Note that Panels~(a) to~(d) are normalized to be on the same dynamic scale (i.e. such that the same value corresponds to each specific gray-scale shade). Similarly, Panels~(e) and~(f) are matched to a dynamic scale over the range of the residuals in both images. Of note is that the residuals for both MRFs and BIFS carry considerable residual structure. This is simply a reflection of the nature of the priors in only representing local characteristics as discussed in Section \ref{discord}

Table~\ref{cfMRF} gives a comparison of mean signal estimates in each tissue type along with overall predictive accuracy based on root-mean-square-error (RMSE) over all pixels in the image. The results are very similar for IG-MRF and BIFS, indicating that BIFS is doing a good job of approximating the results of IG-MRF. Both models shrink estimates of mean tissue levels toward neighboring tissue types due to the overall smoothing effect of the Gaussian pairwise difference prior. This bias effect is largest in GM where the tissue region is narrow and therefore the conditional distributions within GM regions often include neighbors from other tissue types. Note that in this example the smoothing bias is strong enough to increase the RMSE for the Bayesian reconstructions compared with just using the noisy data. This emphasizes the dangers of using Bayesian image analysis with simple priors when the goal is signal estimation.

\begin{table}[!ht]
  \begin{center}
    \caption{Comparison between estimates of mean signal in each tissue type and overall RMSE of reconstructions.}
    \label{cfMRF}
    \begin{tabular}{l|r|r|r|r} 
      & \textbf{True} & \textbf{True + noise} & \textbf{IG-MRF} & \textbf{BIFS}\\
      \hline
      GM & 20.0 & 19.92 & 13.48 & 13.53\\
      WM & 10.0 & 9.99 & 11.10 & 11.08\\
      CSF/out & 0.0 & -0.01 & 0.54 & 0.54\\
      \hline
      RMSE & 0.0 & 2.47 & 2.74 & 2.71 
    \end{tabular}
  \end{center}
\end{table}

\section{The data-driven BIFS Prior (DD-BIFS)} \label{secdatadriven}

The standard process of generating the BIFS prior distribution described in Section \ref{methods} is based on choosing a pair of distributions to be applied as priors at each location in Fourier space (one for the modulus and the other for the argument of the complex value signal) and a set of parameter functions to define how the parameters of the distributions vary over Fourier space. In contrast, for the data-driven approach the parameters of the priors are estimated empirically from a database of high-quality images. 

To estimate the parameters, all of the images in the database are first Fourier transformed, and the distribution parameters for each Fourier space location are estimated from that data. These estimates then specify the parameters for the prior at each Fourier space location, i.e. in combination they form data-driven parameter functions. (Note that when the database is small it may be more beneficial to fit parameter functions to the empirical data rather than use the raw estimates generated separately at each Fourier space location.)

\subsection{Example 5 - Data-driven prior simulation study} \label{simstudy}

To illustrate the DD-BIFS approach we simulated 10,000 256$\times$256 images. Ellipsoid objects were simulated as randomly positioned (and scaled) 2D Gaussian probability density functions (resembling bumps). The number of objects in each image was modeled as Poisson, each object having uniformly distributed random intensity level and standard deviation on each axis and the correlation between the standard deviations on each axis distributed uniformly between~-1 and~0, i.e., so that the process was not isotropic. 

\begin{figure}
\centering
    \begin{subfigure}{\yval \textwidth}
      \centering
       \includegraphics[width=\xval cm,height= \xval cm]{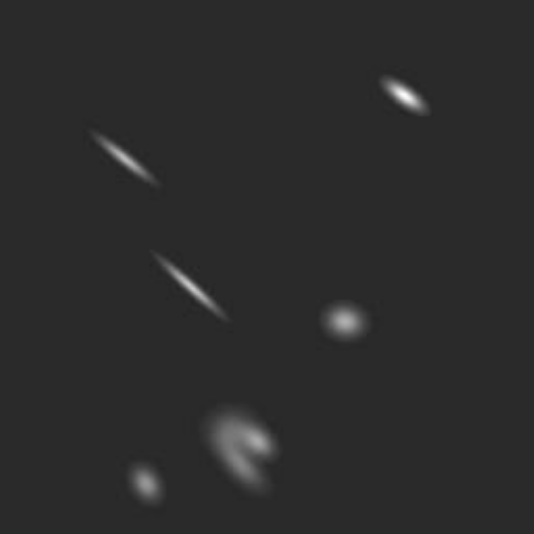}
        \caption{\hspace*{0.0 cm} True signal}\label{gal_a}
    \end{subfigure} %
    \begin{subfigure}{\yval \textwidth}
      \centering
       \includegraphics[width=\xval cm,height=\xval cm]{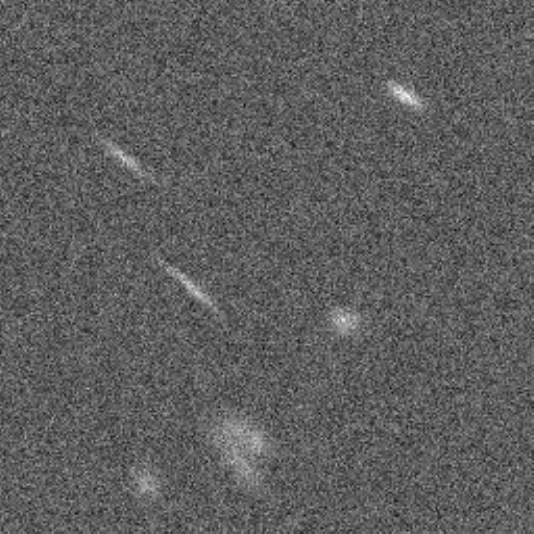}
        \caption{\hspace*{0.0 cm} with noise}\label{gal_b}
    \end{subfigure} %
    \begin{subfigure}{\yval \textwidth}
      \centering
       \includegraphics[width=\xval cm,height=\xval cm]{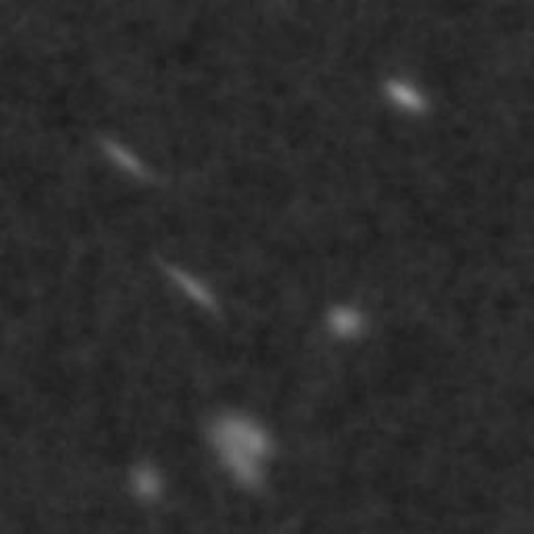}
        \caption{\hspace*{0.0 cm} IG-MRF}\label{gal_c}
    \end{subfigure} %
    
    \vspace*{0.5cm}
    
    \begin{subfigure}{\yval \textwidth}
      \centering
       \includegraphics[width=\xval cm,height=\xval cm]{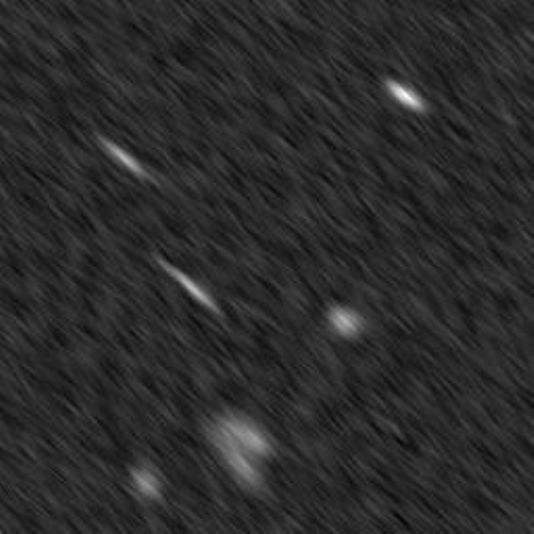}
        \caption{\hspace*{0.0 cm} DD-BIFS $\sim$ 0.1~obs}\label{gal_d}
    \end{subfigure} %
    \begin{subfigure}{\yval \textwidth}
      \centering
       \includegraphics[width=\xval cm,height=\xval cm]{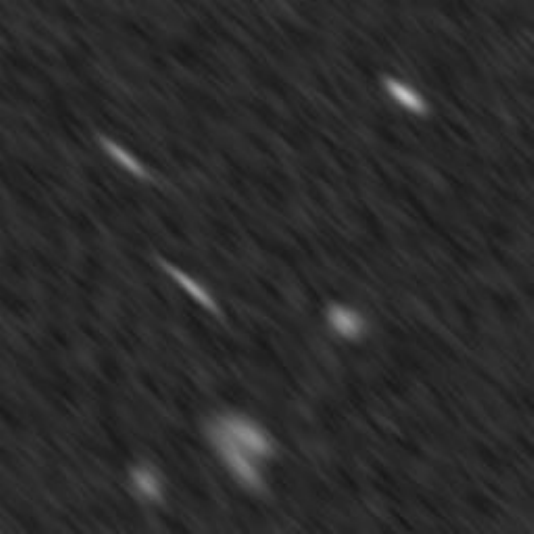}
        \caption{\hspace*{0.0 cm} DD-BIFS $\sim$ 1~obs}\label{gal_e}
    \end{subfigure} %
    \begin{subfigure}{\yval \textwidth}
      \centering
       \includegraphics[width=\xval cm,height=\xval cm]{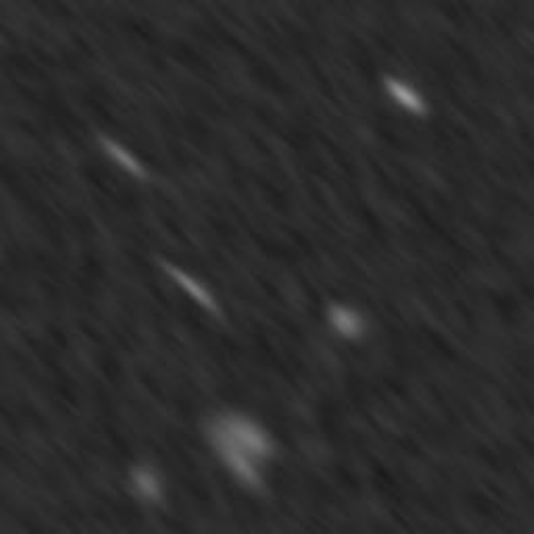}
        \caption{\hspace*{0.0 cm} DD-BIFS $\sim$ 10~obs}\label{gal_f}
    \end{subfigure} %
\caption{Simulation study and reconstruction of anisotropic bump patterns. Panel~(a)~new process realization; (b)~with noise; (c)~first order IG-MRF; (d)~DD-BIFS $\sim$ to 0.1~observation (e)~DD-BIFS $\sim$ 1~observation; (f)~DD-BIFS $\sim$ 10~observations. All panels except for Panel~(b) are normalized to be on the same dynamic scale. \label{bumps} }
\end{figure}

An additional realization (separate to the 10,000 used to build the prior) displayed in Figure~\ref{gal_a} was contaminated with added Gaussian noise (Figure~\ref{gal_b}) and was used as the data to be reconstructed with DD-BIFS.

For this reconstruction we used a deliberately miss-specified prior and likelihood below to allow a simple illustration using conjugate prior forms. At each Fourier space location $k$ we adopt a truncated Gaussian prior for the modulus: $\Modbig(\ft x_k) \sim TN(\mu_k, \tau_k^2, 0, \infty)$, with $\mu_k \ge 0$; a Uniform prior on the circle for the argument: $\Argbig(\ft x_k) \sim U(0,2 \pi)$, a Gaussian noise model for the modulus $\Modbig(\epsilon_k) \sim N(0,\sigma^2)$, and a Uniform noise model for the argument $\Argbig(\epsilon_k) \sim U(0,2 \pi)$,
where $\epsilon_k$ is the complex noise treated as independent across Fourier space locations $k$.  The values of $\mu_k$ and $\sigma_k$ at each Fourier space location are empirically estimated using the approach outlined above.

The global posterior mode is then obtained by generating the (equivalent) posterior mean based on conjugate Bayes for the corresponding non-truncated Gaussian prior and likelihood.
\begin{equation*}
\ft x_{k,\map} = \frac{ \left( \frac{m \mu_k}{\tau_k^2} + \frac{y_k}{\sigma^2} \right) } { \left( \frac{m}{\tau_k^2} + \frac{1}{\sigma^2} \right) }
\end{equation*}
Note that the value of $m$ in the prior is specified by the user and can be considered to represent how many observations we want the weight of the prior to count for in the posterior. 

Panel~(c) of Figure~\ref{bumps} shows an IG-MRF reconstruction of the noisy data. It clearly denoises the image but at the expense of smoothing the objects in the image. Panels~(d) through (f) show DD-BIFS reconstructions where the database prior is given weight equivalent to 0.1, 1.0 and 10.0 observations respectively.  Note that in general for these priors the bumps that are elongated have their length better preserved than in the Gaussian prior case. As the number of observations that the DD-BIFS prior represents increases the features of the true signal begin to diminish and the noise level of the reconstruction is reduced. This is to be expected because in the limit we would effectively be obtaining the MAP estimate based on the prior alone which is an average over 10,000 simulations. Note that this high-level capturing of the bump features occurs despite the independence specification in Fourier space; the BIFS formulation is able to capture the anisotropic characteristics of these features through the empirical parameter function. Note also that the DD-BIFS prior could itself be modified by mixing with a user defined parameter function. For example, if one wanted to diminish the anisotropic features of the background signal the prior could be taken as a function of both the database prior and a denoising parameter function.

\clearpage

\section{Discussion and Conclusion} \label{secconclude}

The BIFS modeling framework provides a new family of Bayesian image analysis models with the capacity to a)~enhance images beyond conventional standard Bayesian image analysis methods; b)~allow straightforward specification and implementation across a wide range of imaging research applications; and c)~enable fast and high-throughput processing. These benefits, along with the inherent properties of resolution invariance and isotropy, make BIFS a powerful tool for the image analysis practitioner.

A clear strength of the BIFS approach is the ease with which one is able to try different prior models. Experimenting with different priors might be considered problematic in other fields, but in Bayesian image analysis we typically know in advance that any model we can specify will be wrong and at best can approximately capture some characteristics from the image; if we simulate from Bayesian image analysis priors we would expect to be waiting an extremely long time before we see a realization that would be representative of an object of interest (e.g. a brain or a car). It is therefore often of benefit to try different models until finding a prior that has the desired impact on the posterior. 

There are many opportunities to expand on the work presented here. BIFS could be applied to spatio-temporal modeling, multi-image analysis, multi-modal medical imaging, color images, 3D images, other spatial basis spaces such as wavelets, multifractal modeling, non-continuous valued MRFs with hidden latent models, etc. We hope that other statisticians, engineers, and computer scientists with an interest in image analysis will begin to explore these potential areas.

\section*{Acknowledgements}

Research reported in this manuscript was supported by National Institute of Biomedical Imaging and Bioengineering of the National Institutes of Health under award number R01EB022055.

Thanks to Hernando Ombao, H{\"a}vard Rue, Konstantinos Bakas, William Braithwaite and Rui Zhang for comments on the manuscript; Chuck McCulloch, John Neuhaus, and Saunak Sen for general advice; Ross Boylan for related Python package programming.

\section*{Appendix: GMRF match and corresponding MAP estimation} \label{gmrfCirc}

An IG-MRF  has a precision matrix $\bm{Q}$ that is block circulant, in which case we can show using the analysis in \citep{rue2005gaussian} that the priors in Fourier space are such that the power at each frequency pair is exponentially distributed.  The proof is straightforward but notationally complex so we present the one-dimensional proof to provide intuition.
 
 If $\bm{Q}$ is a circulant  matrix then we can decompose it as 
 $\bm{Q}=\bm{F\Lambda F}^H$, where $\bm{F}$ is the (discrete) Fourier transform (DFT) matrix, $\bm{F}^H$ is the Hermitian (i.e. conjugate transpose of $\bm{F}$), and $\bm{\Lambda}$ is a diagonal matrix of eigenvalues of $\bm{Q}$.  Now we can compute,
\begin{equation*} 
    \pi(\bm{x}) \propto \exp \left( -\bm{x}^T \bm{Qx} \right) 
    = \exp \left( -\bm{x}^T \bm{F\Lambda F}^H \bm{x} \right) 
    = \exp \left( -\bm{F}^T \bm{x} \bm{\Lambda F}^H \bm{x} \right) 
    = \exp \left( -\bm{F} \bm{x} \bm{\Lambda F}^H \bm{x} \right). \\
\end{equation*} 

Let $f$ be the DFT of $\bm{x}$, $f^\dagger$  the inverse DFT (IDFT) of $\bm{x}$, and $p_k$  the power at frequency $k$, then

\begin{equation*} 
    \pi(\bm{x}) \propto \exp \left( -f \bm{\Lambda} f^\dagger \right)  
    = \exp \left( \sum_k - \lambda_k f_k f^\dagger_k \right) 
    =\exp \left( \sum_k - \lambda_k p_k \right) 
    =\prod_k \exp \left( -\lambda_k p_k \right).
\end{equation*}

 This final term is a product of functions, therefore the distributions at each frequency are independent and each has an exponential distribution.

\begin{equation*}
    \pi(p,\theta) \propto \prod_k \exp \left( - \lambda_k p_k \right).
\end{equation*}

To summarize, this shows that for GMRFs with neighborhood structure specified on the torus, the power spectrum (the square of the signal modulus) is made up of independent exponential random variables. 

\textbf{MAP estimation:} We need to obtain the posterior maximum with this prior where the modulus square follows an exponential distribution coupled with Rician noise, analogous to the approach for Equation~6 where the exponential prior is used directly for the modulus.

Assume that $X \sim \textrm{Exp}(1/m)$ for each point in Fourier space and that $P = \sqrt{X}$. Then for any $\rho \ge 0$;
\begin{equation*}
    \Pr(P \le \rho) = \Pr(\sqrt{X} \le \rho) = \Pr(X \le \rho^2) = 1 - \exp \left(- \frac{\rho^2}{m} \right) \;\;\;\; \rho \ge 0
\end{equation*}
differentiating, we get the pdf for the prior of $\rho$:
\begin{equation*}
   \pi(\rho | m) = \frac{2 \rho}{m} \exp \left( - \frac{\rho^2}{m} \right) \;\;\; \rho \ge 0
\end{equation*}
now apply Bayes' Theorem and take logs to simplify   
\begin{equation*}
    \log  \pi(\rho | r, \sigma, m) = c - \frac{\rho^2}{2\sigma^2} + \log \left(I_0 \left( \frac{r \rho}{\sigma^2} \right) \right) + \log \left( \frac{2 \rho}{m} \right) -\frac{\rho^2}{m}
\end{equation*}
differentiate w.r.t. $\rho$ 
\begin{equation*}
    \dfrac{\mathrm{d}  \log  \pi(\rho | r, \sigma, m)}{\mathrm{d} \rho} = - \frac{\rho}{\sigma^2} + \frac{r I_1 \left( \frac{r \rho}{\sigma^2} \right)}{\sigma^2 I_0 \left( \frac{r \rho}{\sigma^2} \right)} + \frac{1}{\rho} - \frac{2 \rho}{m} 
\end{equation*}
then set equal to zero and solve to get the positive solution for $\rho$ of
\begin{equation*}
    \rho = \frac{rm b(\rho) + \sqrt{\left(b(\rho) r m \right)^2 + 8 \sigma^4 m + 4 (\sigma m)^2}}{4 \sigma^2 + 2 m}.
\end{equation*}

\bibliographystyle{agsm}
\bibliography{bibBIFS} 

\end{document}